\documentclass[aps,prb,twocolumn,amssymb,amsfonts,groupedaddress,showpacs,epsf]{revtex4}
\usepackage{epsfig}
\usepackage{amsmath}
\usepackage{bm}
\usepackage{times}

\begin{document}
\title{Non-Gaussian signatures and collective effects in charge noise affecting a dynamically-decoupled qubit}

\author{Guy Ramon}
\email{gramon@scu.edu}
\affiliation{Department of Physics, Santa Clara University, Santa Clara, CA 95053}

\begin{abstract}

The effects of a collection of classical two-level charge fluctuators on the coherence of a dynamically-decoupled qubit are studied. Distinct dynamics  are found at different qubit working positions. Exact analytical formulae are derived at pure dephasing and approximate solutions are found at the general working position, for weakly- and strongly-coupled fluctuators. Analysis of these solutions, combined with numerical simulations of the multiple random telegraph processes, reveal the scaling of the noise with the number of fluctuators and the number of control pulses, as well as dependence on other parameters of the qubit-fluctuators system. These results can be used to determine potential microscopic models for the charge environment by performing noise spectroscopy.

\end{abstract}

\pacs{03.67.Lx, 03.67.Pp, 73.21.La, 73.23.Hk}

\maketitle

\section{Introduction}

Charge noise is a quintessential decoherence channel in many qubit systems including Josephson junctions,  quantum dots (QDs), and hybrid systems such as electron and nuclear spins in nitrogen-vacancy centers in diamond.\cite{noiseRev} Studies of exchange-coupled electron spin qubits in GaAs QDs, in particular, have shifted their attention from the nuclear to the charge environment, as the important role of the latter has been identified.\cite{Hu,Culcer,Ramon} Recent works include design and implementation of exchange-only three-spin qubits in a triple QD that have better immunity against low-frequency electrical noise,\cite{TayMed} multielectron spin qubits with demonstrated reduced exchange noise,\cite{Higgin} and self-calibrated, optimized pulse sequence\cite{Cerfontaine} and asymmetric double dot geometry\cite{Hiltunen}, both tailored to mitigate charge noise for high-fidelity single-qubit gates in singlet-triplet ($S-T_0$) spin qubits. Charge noise was also shown to cause relaxation in a single electron spin qubit, through the spin-orbit interaction.\cite{Huang}

Despite their key role in limiting the qubit coherence time and gate fidelity, the physical origin of charge fluctuations is still unclear. In superconducting devices, spurious tunneling two-level systems were suggested to reside in the amorphous dielectric covering the circuits, or in the dielectric forming the tunneling barrier in the Josephson junction. Suggested trap mechanisms in semiconductor devices include localized states near gate electrodes inducing leakage currents, charge traps near quantum point contacts, donor centers near the gate surface, and localized switching charges in the doping layer. One of the difficulties in interpreting noise measurements is the inability to distinguish between the various and often system-specific microscopic mechanisms that cause charge fluctuations. While all of the above mechanisms fall into either an Anderson-type model or a tunneling two-level system model, their specific characteristics may lead to very different qubit dynamics, with distinct sensitivity to its working position. It is therefore imperative to establish a theoretical framework that predicts the effects of charge fluctuators on qubit dynamics, including the dependence of noise characteristics on various parameters of the qubit and its charge environment. The theory developed in this paper is aimed to bridge between microscopic modeling of the charge environment and characteristic measurements of the noise spectrum. We refer, for concreteness, to $S-T_0$ spin qubits in GaAs gate-defined double QDs,\cite{Taylor} where charge noise characterization,\cite{Jung,Pioro,Taubert,Petersson} and spectroscopy measurements\cite{Dial} were previously reported, but our results are relevant to any system afflicted with charge noise. For example, we expect that charge noise will play a dominant role in Si, where the hyperfine interaction strength is three orders of magnitude smaller, due to reduced coupling to- and number of nuclear spins, as compared with GaAs.\cite{Maune}

We model the charge environment with a collection of $n_T$ two-level charge fluctuators (TLFs), each characterized by a qubit-TLF coupling strength $v_i$, and asymmetric mean switching rates $\gamma_i^+$ ($\gamma_i^-$) from the upper state to lower state (lower to upper). While quantum treatment of charge fluctuators, coupled to a noninteracting electron reservoir, was carried out before,\cite{Abel} in this study we treat the TLFs as classical sources of random telegraph noise (RTN). This approach, commonly referred to as the spin-fluctuator model, is typically justified when the TLFs couple more strongly to their own environment than to the qubit (over damped fluctuators).\cite{Paladino,GalperinPRL} In their study of the applicability of the classical RTN model, Wold {\it et al.} determined more precisely that the difference between the quantum and classical TLF models depends on the ratio between the qubit-TLF coupling strength and the TLF decoherence rate.\cite{Wold} The latter was defined by the authors as the rate at which the off-diagonal density matrix elements decay in the basis where the equilibrium density matrix is diagonal. In this context, we mention a recent work by Trapani {\it et al.} that quantified the classical to quantum transition using several nonclassicality criteria.\cite{Trapani} Using an open-system approach, where a quantum TLF is coupled to a stochastic classical field, the authors studied non-Markovian effects and demonstrated that time correlations in the classical environment enhance quantum coherence and may induce its collapse and revival.

In our model, the qubit is endowed with control fields that allow us to rotate its state around two perpendicular axes, and perform dynamical decoupling (DD) through sequences of $\pi$ pulses. In the context of $S-T_0$ spin qubits in gate-defined QDs, rotation around the $z$ axis is performed by electrostatic control over the interdot bias that provides highly tunable exchange coupling ($J$).\cite{Petta} Nuclear polarization cycles, in which spin polarization is exchanged between the electrons and the nuclei, generate a hyperfine field gradient across the dots, $\delta h$, that provides qubit rotation around the $x$ axis.\cite{Foletti,BluhmPRL} Other methods to generate local magnetic field gradients were also demonstrated, including on-chip micromagnets.\cite{Brunner} Under these assumptions the system Hamiltonian reads
\begin{equation}
{\cal H} = \left( \Delta \cos \theta  + \sum\limits_{i = 1}^{n_T} {v_i}\xi _i (t)  \right){\sigma _z} + \Delta \sin \theta \sigma _x,
\label{Ham}
\end{equation}
where $\Delta  = \frac{1}{2}\sqrt{J^2 + \delta h^2}$, $\theta  = \tan ^{- 1}\left( \delta h/J \right)$, $\sigma _j$ are the Pauli spin matrices for the pseudospin states $S$ and $T_0$, and $\xi_i (t)=\pm 1$ is a classical noise representing the $i$th random telegraph process, switching between $\pm 1$ with average rates $\gamma_i^\pm$. Throughout the paper, strong (weak) coupling refers to $v_i \gg \gamma_i^\pm$ ($v_i \ll \gamma_i^\pm$) and has nothing to do with the energy scale of the qubit control fields.

The qubit dynamics are crucially dependent on its working position. Figure 1 provides a geometrical representation of the qubit state, where we highlight two commonly used working positions: $J \gg \delta h$ ($\theta \approx 0$), referred to as pure dephasing, and $J \ll \delta h$ ($\theta \rightarrow \pi/2$), referred to as the optimal working point (OP). For $S-T_0$ qubits, stabilizing the field gradient by nuclear state preparation is a relatively long process, and we assume that $\delta h$ is fixed throughout the experiment, thus various working points are accessed by tuning $J$. In this scenario, the pure dephasing regime and OP are realized by positive interdot bias (at or above the avoided singlet crossing) and a large negative bias (where $J$ approaches zero), respectively. At the avoided crossing, the qubit sensitivity to charge fluctuations is heightened and charge noise becomes dominant. This working position is needed for fast $z$-rotations or during two-qubit operations, in order to achieve sizable capacitive coupling between the two double dots. Recognizing the importance of charge noise was a crucial step in the realization of a controlled-PHASE gate between two $S-T_0$ qubits, which became possible by mitigating the noise using a spin echo (SE) pulse along the $x$ axis.\cite{Shulman} While charge noise peaks at pure dephasing, where the exchange interaction is strongest, it is equally important to characterize its effects at or near the OP, where it is envisioned that the qubit will need to maintain its coherence for longer times.
\begin{figure}[tb]
\epsfxsize=0.95\columnwidth
\vspace*{-0.05 cm}
\centerline{\epsffile{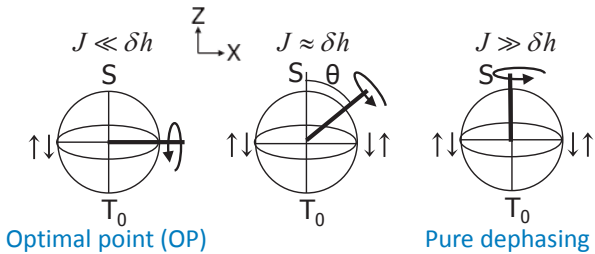}}
\vspace*{-0.18 cm}
\caption{Bloch sphere representation of qubit rotations around an axis determined by the two control fields $J$ and $\delta h$.}
\label{Fig1}
\end{figure}

The application of sequences of control pulses is essential for removing quasi-static noise and extending coherence time. Moreover, DD is valuable in experiments that use the qubit as a noise spectrum analyzer, as the noise sensitivity is peaked at $f \approx 1/\tau$, where $\tau$ is the time interval between control pulses. This allows us to scan the noise spectrum by changing the number of control pulses. Both of these aspects of DD were studied experimentally,\cite{Biercuk,Bylander,Medford,Dial} and theoretically for pure dephasing,\cite{Cywinski} and at the OP.\cite{Cywinski2}

In a previous paper we studied the effects of a single TLF on the qubit coherence under DD sequences of control pulses.\cite{Ramon2} Distinct qubit dynamics were found for different working positions. Specifically, it was shown that at or near the optimal point the qubit state exhibits a multi-exponential decay, with several decay rates whose weights are governed by the TLF parameters. In contrast, at pure dephasing, within the relevant parameter range, the qubit was found to decay with a single rate associated with the TLF switching rate. In the current paper we extend this work to treat TLF ensembles, providing scaling of the noise with ensemble size and analyzing the resulting qubit dynamics at different working positions. Throughout the paper, we employ sequences of $\pi$ pulses around the $y$ axis, which have been realized in several systems, including the $S-T_0$ qubit, albeit with limited fidelity.\cite{Foletti} Although the efficacy of control pulses in mitigating charge noise in the pure dephasing regime is indifferent to their axis within the $x-y$ plane, $\pi_y$ pulses, used here, were shown to be more effective when operating at a general position.\cite{BerFao,Ramon2} In addition, $\pi_y$ pulses should be equally effective in correcting nuclear-induced noise, as compared with the traditionally employed $\pi_z$ pulses. We note that our theory can be applied straightforwardly to analyze any sequence of composite pulses, such as the $XY-4$ self-correcting protocol that has been suggested to be more robust against pulse errors.\cite{Souza}

Since RTN is generally non-Gaussian, one cannot fully characterize it using a noise spectrum, and a correct interpretation of the noise characteristics from qubit signal measurements must consider the non-Gaussian nature of the noise. Here we examine the validity range of the Gaussian approximation, as the number of fluctuators is increased, and its dependence on the number of control pulses. As current experimental efforts are focused on more complicated QD structures, such as two coupled double QDs,\cite{Shulman,Weperen} and three-spin qubits in triple-dots,\cite{Laird,Takakura,Gaudreau} the resulting larger devices are expected to have a noisier charge environment. Understanding how qubit decoherence scales with the number of TLFs is therefore important.

The paper is organized as follows. In Sec.~II we study the case of pure dephasing, where exact analytical results are obtained and their asymptotic behavior is analyzed. In Sec.~III we treat a general working point, where both dephasing and dissipative dynamics are expected, focusing on the optimal point and considering separately weak and strong couplings. A summary of our work is provided in Sec.~IV. In Appendix A we compare our exact results for pure dephasing with the results of a cumulant expansion, allowing us to quantify non-Gaussian behavior, whereas Appendix B details the calculation of the qubit coherence factor at pure dephasing with two fluctuators.

\section{Pure dephasing}

Pure dephasing ($\theta =0$) applies strictly to $\delta h =0$, but our analysis holds also in the vicinity of this point, where $\delta h \ll J$. The results given in this section apply, therefore, to the common experimental scenario where it is much easier to stabilize the magnetic field gradient to a fixed value throughout the measurement. We consider $\delta h=0.1 \mu$eV, which was experimentally demonstrated, \cite{BluhmPRL} and $J=4 \mu$eV, typically measured\cite{Petta,Dial} and calculated\cite{Ramon2} near the singlet anti-crossing.

\subsection{Single Fluctuator}

The case of pure dephasing due to a single RTN source was previously solved for free induction decay (FID) and spin echo (SE),\cite{Laikhtman, Paladino} and was extended to an $N$-pulse periodic DD (PDD) for the case of symmetric fluctuator.\cite{BerFao} Here we reproduce these results using a simple procedure described below and extend them to treat the more commonly used Carr-Purcell-Meiboom-Gill (CPMG) protocol and an asymmetric TLF with different switching rates between its states in each direction ($\gamma^+ \neq \gamma^-$). Working at low temperatures as compared with the TLF level splitting results in longer stays in the lower state. Such asymmetric telegraphic signals were observed in various systems, including tunnel junctions,\cite{Zimmerli} and a Single Electron Tunneling electrometer.\cite{Zorin} Furthermore, our general results for asymmetric TLFs may be relevant in explaining temperature-dependent noise spectroscopy measurements that were recently performed on $S-T_0$ qubits.\cite{Dial} In this context, we mention a recent theoretical work that focused on the temperature-dependence of qubit dephasing, induced by a TLF bath. The authors considered several microscopic mechanisms, including direct tunneling, cotunneling, and coupling of the TLFs to a phonon bath.\cite{Beaudoin}  While none of these mechanisms was fully consistent with the experimental data by itself, some agreement was found by adding an extrinsic dephasing mechanism such as phonon coupling,\cite{Kornich} acting directly on the qubit.

Since both the qubit control field and its coupling to the TLF induce precession about the $z$ axis, the dynamics are fully accounted by a single coordinate. We take the qubit initial state to be along the $x$ axis (equal superposition of its up and down states), and quantify the qubit coherence by calculating its signal decay function, $\chi(t)$, defined as:\cite{Cywinski,RSouza}
\begin{equation}
\chi (t) = \left| \frac{\langle \rho_{+-}(t) \rangle}{\langle \rho_{+-} (0) \rangle} \right| = \langle e^{i \phi} \rangle,
\end{equation}
where $\rho_{+-}$ is the off-diagonal element of the qubit density matrix, and $\phi$ is the random phase accumulated due to the qubit coupling with the TLF. The signal decay is calculated by dividing the probability distribution to partial probabilities, $p(\phi,t)=p_+ (\phi,t)+p_-(\phi,t)$, to accumulate phase $\phi$ while the TLF is in the up or down state:\cite{Bergli,BerFao}
\begin{equation}
\chi (t) = \int d\phi p(\phi,t) e^{i \phi}.
\label{chifirst}
\end{equation}
The corresponding phase factors, $\chi_\pm$, averaged over switching histories, are found by converting the rate equations for $p_\pm(\phi,t)$ to equations for $\chi(t)=\chi_+(t)+\chi_-(t)$ and $\delta \chi(t)=\chi_+(t)-\chi_-(t)$:
\begin{equation}
\left( \!{\begin{array}{*{20}{c}}
{\dot \chi }\\
\dot{ \delta \chi}
\end{array}} \right) = {M_1} \! \left(\! {\begin{array}{*{20}{c}}
{\chi }\\
{\delta \chi }
\end{array}} \right), \hspace{0.3 cm}
{M_1} = \left( \!{\begin{array}{*{20}{c}}
0&{ - iv-\delta \gamma}\\
{ - iv}&{ - 2\gamma }
\end{array}} \right).
\label{M1}
\end{equation}
Here, $\gamma=(\gamma^+ +\gamma^-)/2$ is the average TLF switching rate and $\delta \gamma=\gamma^+ -\gamma^-$ is the switching asymmetry arising from the TLF's level splitting, $\Delta E_t$, according to $\gamma^-/\gamma^+ =e^{-\Delta E_t/k_B T}$. The initial conditions for $\chi$ and $\delta \chi$ are provided by those for the partial probabilities:
\begin{equation}
p_\pm (0)= \frac{\gamma^\mp}{\gamma^+ + \gamma^-}.
\end{equation}
After a $\pi_y$ (or $\pi_x$) pulse, the qubit evolves under
\begin{equation}
{M_2} = \left( {\begin{array}{*{20}{c}}
0&{ iv-\delta \gamma}\\
{  iv}&{ - 2\gamma }
\end{array}} \right).
\end{equation}
Writing $M_2 =L M_1 L$, where $L={\rm diag}(-1,1)$, it is convenient to define the qubit evolution operator as
\begin{equation}
T=\sqrt{e^{M_2 \tau} e^{M_1 \tau}}=Le^{M_1 \tau},
\label{T}
\end{equation}
where $\tau$ is the time interval between pulses, and $T^2$ is the qubit evolution under one full control cycle. For the PDD sequence $\tau_i \equiv \tau = t/(N+1)$ for $1\leq i \leq N+1$, whereas for the CPMG sequence $\tau_i \equiv \tau = t/N$ for $2\leq i \leq N$ and $\tau_1=\tau_{N+1}=\tau/2$. Here and throughout the paper, $N$ is the number of control pulses.
Assuming the qubit initially lies on the equator, its signal decay after $N$ pulses is calculated from
\begin{equation}
\chi_{PDD} (t)= \left( \! \begin{array}{cc} 1& 0 \end{array} \! \right)
T^{N+1}
\left( \! \begin{array}{c} 1 \\ -\frac{\delta \gamma}{2 \gamma} \end{array} \! \right)
\end{equation}
for a PDD sequence and
\begin{equation}
\chi_{CP} (t)= \left( \! \begin{array}{cc} 1& 0 \end{array} \! \right)
L^{N-1} T_{1/2} T^{N-1} T_{1/2}
\left( \! \begin{array}{c} 1 \\ -\frac{\delta \gamma}{2 \gamma} \end{array} \! \right)
\label{CP}
\end{equation}
for a CPMG sequence, where $T_{1/2}$ is the evolution operator during the first and last, $\tau/2$, pulse intervals. We find $\chi_{PDD} (t)$ and $\chi_{CP}(t)$ by diagonalizing $T$:
\begin{equation}
S T S^{-1} = \left( \! \begin{array}{cc} \lambda_- & 0 \\ 0 & \lambda_+ \end{array} \! \right),
\label{STS}
\end{equation}
where the eigenvalues of $T$ are found to be
\begin{eqnarray}
\lambda_\pm \!&\!=\!&\! \sqrt{A} \pm \sqrt{A +|\mu|^2}, \nonumber \\
A \!&\! \equiv \!&\! (1+\mu_I^2) \sinh^2 \gamma \mu_R \tau +(1-\mu_R^2) \sin^2 \gamma \mu_I \tau,
\label{lasym}
\end{eqnarray}
and the columns of $S^{-1}$ are the corresponding eigenvectors. In Eq.~(\ref{lasym}), $\mu_R$ and $\mu_I$ are the real and imaginary parts of
\begin{equation}
\mu= \sqrt{1-\left(\frac{v}{\gamma}\right)^2+\frac{2iv \delta \gamma}{2\gamma^2}}.
\end{equation}
The solutions for the qubit decay under PDD and CPMG sequences are found as:
\begin{eqnarray}
\!\!&\!\!\chi_{PDD}\!\!\!\!&\!\ (t) \! = \frac{e^{-\gamma t}}{|\mu|^{N+1}} \left\{ \frac{\lambda_+^{N+1} -\lambda_-^{N+1}}{\lambda_+^2 -\lambda_-^2} \left[ (1+\mu_I^2) \mu_R \times \right. \right. \nonumber \\ && \left. \left.  \sinh 2\gamma \mu_R \tau +(1-\mu_R^2) \mu_I \sin 2 \gamma \mu_I \tau \right]+ \! \right. \nonumber \\ && \left. \frac{\lambda_+^{N+1} +\lambda_-^{N+1}}{2\left(\lambda_+^2 +\lambda_-^2 \right)} \right\},
\label{depPDD}
\end{eqnarray}
and
\begin{eqnarray}
\!\!&\!\!\chi_{CP}\!\!\!\!&\!\ (t) \! = \frac{e^{-\gamma t}}{|\mu|^{N+1}} \left\{ \frac{\lambda_+^N -\lambda_-^N}{\lambda_+ -\lambda_-} \left[ (1+\mu_I^2) \cosh \gamma \mu_R \tau - \right. \right. \nonumber \\ && \left. \left. \!\!\!\!\!\!\!\!\!\! (1\!-\!\mu_R^2) \cos \gamma \mu_I \tau \right] \! + \frac{\lambda_+^N \! +\lambda_-^N}{\lambda_+ \!+\lambda_-} \left[ (1\! +\mu_I^2) \mu_R \sinh \gamma \mu_R \tau + \right. \right. \nonumber \\ && \left. \left. \!\!\!\!\!\!\!\!\!\! (1\!-\!\mu_R^2) \mu_I \sin \gamma \mu_I \tau \right] \right\}.
\label{depCP}
\end{eqnarray}

We note that Eqs.~(\ref{depPDD}) and (\ref{depCP}) coincide for $N=1$, when time intervals are taken as prescribed ($\tau =t/2$ for PDD and $\tau=t$ for CPMG). In this case both equations reproduce the previously reported spin-echo (SE) decay:\cite{Ramon2}
\begin{eqnarray}
\chi_{\rm SE} (t)\!&\!=\!&\! \frac{e^{-\gamma t}}{2|\mu|^2} \left[ (\mu_I^2+1) \sum_\pm (1\pm \mu_R) e^{\pm \gamma \mu_R t} \right. \nonumber \\
\!&\!+\!&\! \left. (\mu_R^2-1) \sum_\pm (1 \pm i\mu_I) e^{\pm i\gamma \mu_I t} \right]. \label{chise}
\end{eqnarray}

For a symmetric fluctuator, $\gamma^+ = \gamma^- =\gamma$, realized with zero TLF level-splitting, or at high temperature, $\mu$ is either real or pure imaginary. The above results simplify and
for either real or imaginary $\mu$ (corresponding to $v/\gamma < 1$ or $v/\gamma >1$, respectively) they reduce to:
\begin{eqnarray}
\chi_{PDD}^{\rm sym} (t) \!&\!=\!&\! \frac{e^{-\gamma t}}{2\mu^{N+1}} \left[  \frac{ \mu \cosh \gamma \mu \tau}{ \sqrt{\sinh^2 \gamma \mu \tau +\mu^2}} \left(\lambda_+^{N+1} \!-\!\lambda_-^{N+1} \right) + \right. \nonumber \\ && \left. \left( \lambda_+^{N+1} +\lambda_-^{N+1} \right) \right],
\label{depPDDs}
\end{eqnarray}
previously reported,\cite{BerFao} and
\begin{eqnarray}
\chi_{CP}^{\rm sym} (t) \!&\!=\!&\! \frac{e^{-\gamma t}}{2\mu^N} \left[  \frac{\cosh \gamma \mu \tau-v^2/\gamma^2}{\mu \sqrt{\sinh^2 \gamma \mu \tau +\mu^2}} \left(\lambda_+^N -\lambda_-^N \right) + \right. \nonumber \\ && \left. \left( \lambda_+^N +\lambda_-^N \right) \right],
\label{depCPs}
\end{eqnarray}
with
\begin{equation}
\lambda_\pm^{\rm sym} = \sinh \gamma \mu \tau \pm \sqrt{\sinh^2 \gamma \mu \tau +\mu^2}.
\label{lsym}
\end{equation}

The above exact results should be compared with the widely used Gaussian approximation, in which the signal decay for an arbitrary control sequence can be expressed by the spectral density of the noise, $S(\omega)$, and a so called filter function, $F(\omega t)$, as:\cite{Cywinski}
\begin{eqnarray}
\chi (t) \!&\!=\!&\! e^{-\langle \Phi^2 (t) \rangle/2 } \nonumber \\
\langle\Phi^2 (t) \rangle \!&\!=\!&\! \int_{-\infty}^\infty S(\omega) \frac{F(\omega t)}{\omega^2} d\omega.
\label{Gauss}
\end{eqnarray}
The power spectrum of a single RTN source reads:\cite{Machlup}
\begin{equation}
S(\omega) =\frac{v^2}{2\pi} \frac{\gamma^+ \gamma^-}{\gamma} \frac{1}{\omega^2+(2\gamma)^2},
\end{equation}
where $\gamma=(\gamma^++\gamma^-)/2$ is the average TLF switching rate. Focusing on the CPMG sequence, the filter function is:\cite{Cywinski}
\begin{equation}
F_{CP}(z)=32\frac{ \sin^4 \frac{z}{4N}}{\cos^2 \frac{z}{2N}} \left( \cos z \pm 1\right),
\end{equation}
and we find a closed-form expression for the second moment:
\begin{eqnarray}
\langle \Phi^2_{CP}(t) \rangle \!&\!=\!&\! \frac{v^2 \gamma^+ \gamma^-}{\gamma^4} \left[ \gamma t-N \tanh \frac{\gamma t}{N} - \right. \nonumber \\
&& \left. \frac{1}{2}\left( 1 \pm e^{-2\gamma t} \right) \left(1-{\rm sech} \frac{\gamma t}{N}\right)^2 \right],
\label{GaussCP}
\end{eqnarray}
where the upper (lower) sign in the above two equations corresponds to odd (even) number of pulses. Eq.~(\ref{GaussCP}) reduces to the Gaussian approximation result previously reported for SE (setting $N=1$),\cite{GalperinPRL, RSouza} as well as to the short and long time asymptotic behavior for large number of pulses, given by Cywi\'{n}ski {\it et.~al}.\cite{Cywinski} The Gaussian result can be seen as the first nonvanishing term in a cumulant expansion, whose {\it n}th term is proportional to $(v/\gamma)^n$, thus our exact result, Eqs.~(\ref{depCPs}) and (\ref{lsym}), should converge to the Gaussian expression in the weak coupling limit. In Appendix A we show that explicitly by calculating the next nonvanishing (fourth) cumulant. These higher-order noise correlators quantify the non-Gaussian contributions to qubit dephasing under a given pulse sequence.

\begin{figure}[tb]
\epsfxsize=0.9\columnwidth
\vspace*{-0.0 cm}
\centerline{\epsffile{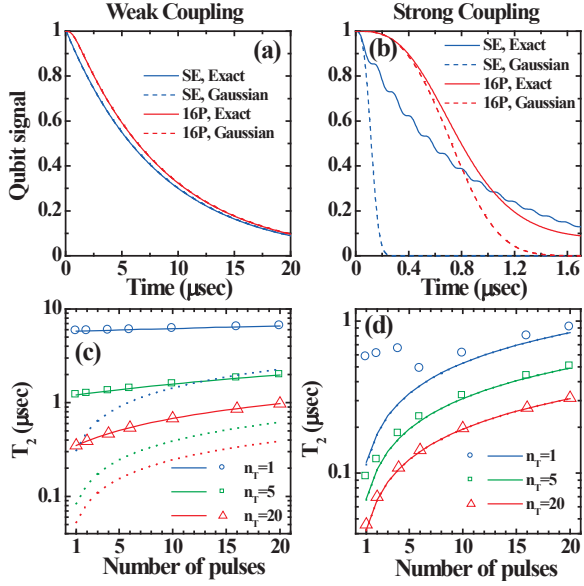}}
\vspace*{-0.1 cm}
\caption{(color online) Qubit signal decay vs.~time for SE and 16-pulse CPMG at pure dephasing, calculated using the exact solution, Eq.~(\ref{depCPs}), (solid lines) and the Gaussian approximation, Eq.~(\ref{GaussCP}) (dashed lines). (a) Single weakly-coupled TLF; (b) single strongly-coupled TLF. Figures (c) and (d) depict dephasing times vs.~the number of control pulses for 1, 5, and 20 identical TLFs at weak and strong coupling, respectively. Symbols (solid lines) correspond to the exact solution (Gaussian approximation). The short time limit, Eq.~(\ref{short}), is also shown by dotted lines (notice that the latter completely coincides with the Gaussian approximation in the strong coupling regime). TLF parameters are $\gamma=0.1 \mu$eV, $v=0.01\mu$eV for Figs.~(a) and (c), and $\gamma=5 $neV, $v=0.2\mu$eV for Figs.~(b) and (d).}
\label{Fig2}
\end{figure}
Figs.~\ref{Fig2}(a) and (b) depict qubit dephasing due to a single weakly and strongly coupled TLF, calculated using the exact result, Eq.~(\ref{depCPs}), and the Gaussian approximation, Eq.~(\ref{GaussCP}), for SE and 16-pulse CPMG. The Gaussian result holds well throughout the entire decay timescale for weak coupling [dashed and solid lines coincide in Fig.~\ref{Fig2}(a)], whereas pronounced  non-Gaussian behavior develops in the strong coupling case, dominating the qubit signal. As the number of pulses increases, the  deviations from Gaussian behavior are pushed to longer times, where their effect on the qubit decay becomes less significant.

\subsection{Many Fluctuators}

Charge fluctuators generate qubit dephasing by shifting its energy levels, thereby inducing random phase-kicks to its two states.\cite{BerglinT} Denoting the sum of the contributions from all TLFs as $v_{n_T}(t)=\sum_{i=1}^{n_T} v_i \xi_i (t)$, the eigenvalues of the Hamiltonian, Eq.~(\ref{Ham}), read \begin{equation}
\Lambda_\pm (t)=\pm \sqrt{\Delta^2+ v_{n_T}^2(t)+2\Delta v_{n_T}(t) \cos \theta}.
\label{Hameigs}
\end{equation}
At pure dephasing ($\theta=0$), the coupling of the qubit to the TLFs is linear, as seen by the linearity of the eigenvalues in $v_{n_T}$. As a result, the extension of the single-TLF results to any number of fluctuators is done straightforwardly by multiplying all coherence factors:
\begin{equation}
\chi (t)=\prod_{i=1}^{n_T} \chi_i (t).
\label{chint}
\end{equation}
It is instructive to obtain this result by extending the single TLF formulation, outlined in the previous section to two or more TLFs. This is done in Appendix B, which also serves to illustrate our approach to solve the $n_T>1$ problem that we later apply to the general working point, where the effects of the fluctuators do not simply factor out. In the Gaussian limit, Eq.~(\ref{chint}) leads to a qubit decay rate that is a sum over $n_T$ decay rates, weighted by TLF parameter distribution.\cite{Galperin}

Our exact results allow us to extend previous studies of free induction and spin-echo,\cite{GalperinB} by analyzing the general conditions for the onset of non-Gaussian qubit dynamics. Focusing on symmetric TLFs under CPMG, we consider various asymptotic limits of Eq.~(\ref{chint}). First, at short time, $\gamma_i, v_i \ll t^{-1}$:
\begin{equation}
- \ln {\chi _{CP}} \longrightarrow \frac{t^3}{6N^2} \sum\limits_{i = 1}^{n_T} {{\gamma _i}v_i^2},
\label{short}
\end{equation}
suggesting similar time- and $N$-dependence as that of a Gaussian noise with a soft ($\omega^2$) cutoff. Similarly, we find the weak coupling ($v_i, t^{-1} \ll \gamma_i$) and strong coupling ($\gamma_i, t^{-1} \ll v_i$) asymptotic behaviors as:
\begin{equation}
- \ln \chi _{CP} \longrightarrow \left\{ {\begin{array}{*{20}{l}}
{\frac{t}{2}\sum\limits_{i = 1}^{{n_T}} {\frac{{v_i^2}}{{{\gamma _i}}}} ,}&{\frac{{{\gamma _i}t}}{N} \gg 1}\\
{\frac{{{t^3}}}{{6{N^2}}}\sum\limits_{i = 1}^{{n_T}} {{\gamma _i}v_i^2} ,}&{\frac{{{\gamma _i}t}}{N} \ll 1}
\end{array}} \right.
\label{weak-asym}
\end{equation}
and
\begin{equation}
-\ln \chi _{CP} \longrightarrow  \left\{ {\begin{array}{*{20}{l}}
{\sum\limits_{i = 1}^{{n_T}} \left({{\gamma _i}t - \frac{{N{\gamma _i}}}{{{v_i}}}\sin \frac{{{v_i}t}}{N}} \right) ,}&{\frac{{{v_i}t}}{N} \gg 1}\\
{\frac{{{t^3}}}{{6{N^2}}}\sum\limits_{i = 1}^{{n_T}} {{\gamma _i}v_i^2} ,}&{\frac{{{v_i}t}}{N} \ll 1}
\end{array}} \right.
\label{strong-asym}
\end{equation}
respectively. These asymptotes elucidate the interplay between TLF parameters, ensemble size, and number of control pulses, in determining the qubit dephasing dynamics. First we observe that both the short- and long-time limits for the weak coupling case can be obtained directly from the Gaussian result, Eq.~(\ref{GaussCP}), reaffirming the validity of the Gaussian approximation for weakly coupled TLFs. In contrast, in the strong-coupling case, only the short-time limit converges to the short-time Gaussian result, demonstrating the onset of non-Gaussian effects at longer times [see Fig.~\ref{Fig2}(b)].

We confirm as expected that the noise becomes Gaussian with sufficiently large number of control pulses.\cite{Ramon2,Cywinski2} More precisely, the Gaussian limit is reached when the pulses are sufficiently frequent, i.e., when $\tau=t/N \ll \gamma^{-1}, v^{-1}$, corresponding to the short time asymptotic. Lastly, we expect that the Gaussian limit will be reached with fewer control pulses as the number of TLFs increases, as it is known that $1/f$ (Gaussian) noise can be generated from a large ensemble of TLFs with a uniform distribution of $\log \gamma_i$.\cite{Schriefl, Bergli} This is demonstrated in Figs.~\ref{Fig2}(c) and (d) for identical TLFs, where we depict qubit dephasing time, $T_2$, defined as signal drop time to $50\%$. For a single strongly-coupled TLF, deviations from Gaussian behavior are observed for any reasonable number of control pulses, whereas Gaussianity is completely restored with 20 TLFs. In the weak coupling regime, where the Gaussian result holds for any number of pulses and TLFs for the chosen parameters, the TLFs switch many times between control pulses and we are in the motional narrowing regime, where the long-time limit holds (compare with the short time limit result depicted by dotted lines). Here, increasing $N$ has little effect on the qubit coherence, up to unrealistic number of pulses [the weak-coupling long-time limit given in Eq.~(\ref{weak-asym}) is strictly independent of $N$, but subleading contributions have a mild $N$-dependence, as seen in Fig.~\ref{Fig2}(c)]. Alternatively, increased number of TLFs will result in a shorter timescale for the qubit decay and a departure from the motional narrowing regime, accompanied with a great benefit from increasing the number of the control pulses. For the chosen parameters, we find that the short-time limit is reached for 200 TLFs with $N=20$.

\begin{figure}[tb]
\epsfxsize=0.9\columnwidth
\vspace*{-0.0 cm}
\centerline{\epsffile{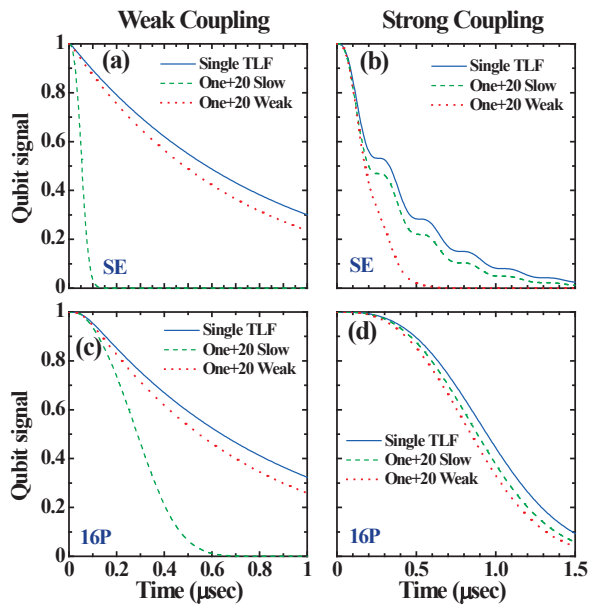}}
\vspace*{-0.0 cm}
\caption{(color online) Qubit dephasing due to a single TLF and 20 additional background fluctuators. In each case we consider identical slow ($\gamma'=0.01\gamma, v'=v$) and weak ($\gamma'=\gamma, v'=0.1v$) background TLFs. (a), (c) Main TLF is weakly-coupled ($\gamma=1 \mu$eV, $v=0.1\mu$eV) for SE and 16-pulse CPMG; (b), (d) Main TLF is strongly-coupled ($\gamma=0.01\mu$eV, $v=0.1\mu$eV) for SE and 16-pulse CPMG.}
\label{Fig3}
\end{figure}

To further demonstrate the implications of these results, we consider in Figure \ref{Fig3} the impact of background TLFs added to a single weakly- or strongly-coupled TLF. We add 20 identical TLFs, which are either hundred time slower ($\gamma'=0.01\gamma$) or ten times weaker ($v'=0.1v$) than the main TLF, such that the additional background contribution is comparable in the short time limit, where the qubit decay rate is proportional to $\gamma v^2$. In the following, we refer to slow and weak background TLFs, as they relate to the main TLF parameters, $\gamma$ and $v$. In Figs.~\ref{Fig3}(a) and (c) we show the effect of adding 20 slow (green-dashed lines) or weak (red-dotted lines) TLFs to a single weakly-coupled TLF, under SE and 16-pulse CPMG, respectively. Without background TLFs, we are in the long-time regime, where increased number of control pulses has little effect [see Fig.~\ref{Fig2}(c)]. In this case the contribution of the additional weak TLFs to qubit decay is relatively small, since they are also motional narrowed, and the resulting combined decay rate increases by only $20\%$ ($v^2+20v'^2=1.2v^2$). In contrast, the slow background TLFs have a short-time dynamics, resulting in a dominant contribution to qubit dephasing. The $1/N^2$ dependence of the decay rate in the short-time limit reduces the effect of the weak background TLFs as $N$ increases [compare the green-dashed lines in Figs.~\ref{Fig3}(a) and (c)].

Turning to the case of a strongly-coupled TLF, depicted in Figs.~\ref{Fig3}(b) and (d), the importance of slow and weak background TLFs is reversed. Here, the timescale of the qubit decay is set by $vt/N$ so that for SE, the contribution of the main TLF can be approximated by the long-time asymptote, with its characteristic plateaus [see Eq.~(\ref{strong-asym})]. The additional contribution from slow TLFs is also adequately given by the long-time result, generating a small effect. Weak background TLFs, on the other hand, exhibit a short-time behavior inducing a large effect. Again, the short-time $1/N^2$ dependence of the decay rate results in a smaller effect of the weak background TLFs with increased number of pulses [compare red-dotted lines in Figs.~\ref{Fig3}(b) and (d)], thus we conclude that at pure dephasing the effects of both slow and weak background TLFs is small when the main TLF is strongly-coupled to the qubit, and a sufficiently large number of control pulses is being used.

\section{General Working Point}

Unlike the case of pure dephasing, there is no exact analytical result for qubit decoherence due to RTN at the general working point, where the qubit is expected to undergo both dephasing and dissipative dynamics. Previous works studied the weak and strong coupling limits of a single TLF at the OP,\cite{BerFao} and at a general working position.\cite{Ramon2} In this section we extend these studies to treat any number of TLFs. The resulting coupled equations are generally solved numerically, but we are able to obtain analytical results for the weak and strong coupling regimes. Below we explain how to extend the single TLF case to an ensemble of TLFs by outlining the necessary steps for the case of two TLFs. The two-TLF solution is then generalized to any number of fluctuators. In order to reduce clutter, we present and solve the equations for the case of symmetric TLFs, but our simulations implement the more general formulation. The procedure is described in greater detail for the single TLF case in Ref.~\onlinecite{Ramon2}.

Considering the Hamiltonian, Eq.~(\ref{Ham}), with $n_T=2$, we can write it as ${\cal H}(t) = {\bf B}_{mn} \cdot \bm{\sigma}$, where $\bm{ \sigma} $ is the vector of Pauli matrices, and the time dependence is implied by evolution under any of the four effective fields:
\begin{eqnarray}
{\bf B}_{mn} \!&\!=\!&\!(\Delta \sin \theta, 0, \Delta \cos \theta +mv_1+nv_2) \nonumber \\
m,n \!&\!=\!&\! \{+,-\}
\label{Bpm}
\end{eqnarray}
corresponding to the four possible two-fluctuator states. At any given time, the qubit Bloch vector rotates around one of these fields and can thus reach any point on the Bloch sphere. We denote $p({\bf r},t)$  the probability to reach point ${\bf r}=(x,y,z)$ on the Bloch sphere at time $t$, and divide it into four partial probabilities, $p({\bf r},t)=\sum_{m,n} p_{mn} ({\bf r},t)$, to reach the point ${\bf r}$ when the two TLFs are in states $m$ and $n$. In analogy to Eq.~(\ref{ppp}) in Appendix B for the pure dephasing case, we have

\begin{eqnarray}
p_{++}(\!\!&\!{\bf r}\!&\!\! ,t +\tau )\! = (1\!-\!\gamma _1 \tau )(1\!-\! \gamma _2\tau )p_{++}(U_{++}^{-1}{\bf r},t) + \nonumber \\ && \!\!\!\!\!\!\!\!\!\!\!\! (1 \!-\! \gamma _1\tau )(\gamma _2\tau) p_{+-}(U_{+-}^{-1}{\bf r} ,t) + (\gamma _1 \tau)(1 \!-\! \gamma _2\tau ) \times \nonumber \\ && \!\!\!\!\!\!\!\!\!\!\!\!  p_{-+}(U_{-+}^{-1} {\bf r},t) \! + (\gamma _1 \tau) (\gamma _2 \tau) p_{--}(U_{--}^{-1}{\bf r} ,t),
\end{eqnarray}
and similar equations for the other three partial probabilities. Here, $U_{mn}=e^{\tau {\bf B}_{mn} \cdot {\bf R}}$ rotates the qubit around the ${\bf B}_{mn}$ field, where ${\bf R}=(R_x, R_y, R_z)$ are $3D$ rotation matrices,\cite{BerFao} such that:
\[
({\bf B}_{mn}\! \cdot {\bf R} ){\bf r}\!=\! \Delta \sin \theta (0, \!-\!z,y)\!+ \!(\Delta \cos \theta \!+ mv_1\!+nv_2)(\!-y,x,0).
\]
Taking an infinitesimal time step and keeping only linear terms in $\tau$, we find the following rate equations for $p_{mn}$:
\begin{eqnarray}
{\dot p_{++}} \!\!&\!=\!&\!\! -(\gamma_1\!+\! \gamma_2) p_{++}\!+\! \gamma_1 p_{-+} \!+\!\gamma_2 p_{+-} \!-\! \nonumber \\ && \!\!\!\!\!\!\!\!\!\!\!\! \left[ \Delta \sin \theta (y \partial _z \! - \! z\partial _y) \!+\! \left( \Delta \cos \theta \!+\! v_1 \! +\! v_2 \right) \! (x\partial_y \!-\! y\partial_x) \right]p_{++} \nonumber \\
{\dot p_{+-}} \!\!&\!=\!&\!\! -(\gamma_1\!+\! \gamma_2) p_{+-}\!+\! \gamma_1 p_{--} \!+\!\gamma_2 p_{++} \!-\! \nonumber \\ && \!\!\!\!\!\!\!\!\!\!\!\! \left[ \Delta \sin \theta (y \partial _z \! - \! z\partial _y) \!+\! \left( \Delta \cos \theta \!+\! v_1 \! -\! v_2 \right) \! (x\partial_y \!-\! y\partial_x) \right]p_{+-}  \nonumber \\
{\dot p_{-+}} \!\!&\!=\!&\!\! -(\gamma_1\!+\! \gamma_2) p_{-+}\!+\! \gamma_1 p_{++} \!+\!\gamma_2 p_{--} \!-\! \nonumber \\ && \!\!\!\!\!\!\!\!\!\!\!\! \left[ \Delta \sin \theta (y \partial _z \! - \! z\partial _y) \!+\! \left( \Delta \cos \theta \!-\! v_1 \! +\! v_2 \right) \! (x\partial_y \!-\! y\partial_x) \right]p_{-+}  \nonumber \\
{\dot p_{--}} \!\!&\!=\!&\!\! -(\gamma_1\!+\! \gamma_2) p_{--}\!+\! \gamma_1 p_{+-} \!+\!\gamma_2 p_{-+} \!-\! \label{peq} \\ && \!\!\!\!\!\!\!\!\!\!\!\! \left[ \Delta \sin \theta (y \partial _z \! - \! z\partial _y) \!+\! \left( \Delta \cos \theta \!-\! v_1 \! -\! v_2 \right) \! (x\partial_y \!-\! y\partial_x) \right]p_{--}. \nonumber
\end{eqnarray}
Next we translate these rate equations to a set of 12 coupled equations for the partial Bloch vector components, evolving under the fields ${\bf B}_{mn}$, defined as:
\begin{equation}
{\bf r}_{mn}=\int d{\bf r} p_{mn} ({\bf r},t) {\bf r}.
\end{equation}
In analogy with the concept of partial probabilities, defined below Eq.~(\ref{chifirst}), ${\bf r}_{mn}$ indicate the contributions to the coordinates of the qubit Bloch vector, coming from the four two-TLF states. Finally, following our treatment of the pure dephasing case [see Eq.~(\ref{chi14})], we construct combinations of ${\bf r}_{mn}$:
\begin{eqnarray}
{\bf r} (t)\!&\!=\!&\! {\bf r}_{++}(t)+{\bf r}_{+-}(t)+{\bf r}_{-+}(t)+{\bf r}_{--}(t) \nonumber \\
{\bf r}_1(t)\!&\!=\!&\! {\bf r}_{++}(t)+{\bf r}_{+-}(t)-{\bf r}_{-+}(t)-{\bf r}_{--}(t) \nonumber\\
{\bf r}_2(t)\!&\!=\!&\!{\bf r}_{++}(t)-{\bf r}_{+-}(t)+{\bf r}_{-+}(t)-{\bf r}_{--}(t)\nonumber \\
{\bf r}_3(t)\!&\!=\!&\! {\bf r}_{++}(t)-{\bf r}_{+-}(t)-{\bf r}_{-+}(t)+{\bf r}_{--}(t),
\label{r14}
\end{eqnarray}
where the first vector, ${\bf r}(t)$, is the actual (full) Bloch vector. These particular combinations are chosen so that the resulting set of equations can be easily decoupled into two blocks when $\theta =0$. Using Eqs.~(\ref{peq})-(\ref{r14}), we find the following set of coupled equations for $({\bf r}, {\bf r}_1, {\bf r}_2, {\bf r}_3)$:
\begin{eqnarray}
&& \!\!\!\!\!\! \left[
\begin{array}{lll} \dot{x} \!=\! -\Delta \cos \theta y-v_1 y_1-v_2 y_2 \\
\dot{y}\!=\! \Delta \cos \theta x +v_1x_1+v_2 x_2-\Delta \sin \theta z \\
\dot{z}\!=\! \Delta \sin \theta y
\end{array}
\right. \nonumber \\
&& \!\!\!\!\!\! \left[
\begin{array}{lll} \dot{x}_1 \!=\! -2\gamma_1 x_1-\Delta \cos \theta y-v_1 y-v_2 y_3 \\
\dot{y}_1\!=\! -2\gamma_1 y_1+\Delta \cos \theta x_1 +v_1 x +v_2 x_3-\Delta \sin \theta z_1 \\
\dot{z}_1\!=\! -2\gamma_1 z_1 +\Delta \sin \theta y_1
\end{array}
\right. \nonumber \\
&& \!\!\!\!\!\! \left[
\begin{array}{lll} \dot{x}_2 \!=\! -2\gamma_2 x_2-\Delta \cos \theta y_2-v_1 y_3-v_2 y \\
\dot{y}_2\!=\! -2\gamma_2 y_2+\Delta \cos \theta x_2 +v_1 x_3 +v_2 x-\Delta \sin \theta z_2 \\
\dot{z}_2\!=\! -2\gamma_2 z_2 +\Delta \sin \theta y_2
\end{array}
\right. \label{r14full} \\
&& \!\!\!\!\!\! \left[
\begin{array}{lll} \dot{x}_3 \!=\! -2(\gamma_1\!+\!\gamma_2) x_3\!-\Delta \cos \theta y_3\!-v_1 y_2\!-v_2 y_1 \\
\dot{y}_3\!=\! -2(\gamma_1\!+\!\gamma_2) y_3\!+\Delta \cos \theta x_3 \!+v_1 x_2 \!+v_2 x_1 \!-\!\Delta \sin \theta z_3 \\
\dot{z}_3\!=\!-2(\gamma_1\!+\! \gamma_2) z_3 +\Delta \sin \theta y_3
\end{array}
\right. \nonumber
\end{eqnarray}

Eqs.~(\ref{r14full}) are conveniently written in a matrix form:
\[
{\dot{\bf k}} =M_1 {\bf k},
\]
with the $12-D$ vector:
\begin{equation}
{\bf k}=(y,z,x_1,x_2,y_3,z_3 ; x,x_3,y_1,y_2,z_1,z_2).
\label{k}
\end{equation}
After a control $\pi_y$ pulse, the qubit evolves with $M_2$, found by substituting $\Delta \rightarrow -\Delta, v_i \rightarrow -v_i$ in $M_1$, which can be written as
\[
M_2=LM_1 L
\]
where $L={\rm diag} (-1,1,1,1,-1,1;1,1,-1,-1,1,1)$. The solution for the time-dependent Bloch vector components is found for either PDD or CPMG protocols, by calculating the eigenvalues of the evolution operator for one full control cycle, $T=L e^{M_1 \tau}$, and their respective weights for a given DD sequence, as detailed in section II for the pure dephasing single-TLF case.

At the optimal point, the above set of 12 equations decouples to two blocks separated by the semicolon in Eq.~(\ref{k}), hence the particular coordinate ordering. Here and throughout the rest of the paper, we assume the qubit is initially prepared along the $z$ axis. At the OP, this means there is no $x$ dynamics and we only need to consider the first block, ${\bf k}^{OP}=(y,z,x_1,x_2,y_3,z_3)$, for which the dynamics is determined by
\begin{equation}
M_1^{OP}\!=\!\!\left(\!\!\!\! \begin{array}{cccccc} 0 \!&\!\!\! -\Delta \!&\! v_1 \!&\! v_2 \!&\! 0 \!&\! 0 \\
\Delta \!&\! 0 \!&\! 0 \!&\! 0 \!&\! 0 \!&\! 0 \\
-v_1 \!&\! 0 \!&\!\! -2 \gamma_1 \!&\! 0 \!&\! -v_2 \!&\! 0 \\
-v_2 \!&\! 0 \!&\! 0 \!&\!\! -2 \gamma_2 \!&\! -v_1 \!&\! 0 \\
0 \!&\! 0 \!&\! v_2 \!&\! v_1 \!&\!\!\! -2 (\gamma_1\!+\! \gamma_2)\! \!&\! -\Delta \\
0 \!&\! 0 \!&\! 0 \!&\! 0 \!&\! \Delta \!&\!\!\! -2(\gamma_1\!+\!\gamma_2) \end{array} \!\!\!\right)\!\!. \label{M1OP}
\end{equation}

The diagonalization of the evolution operator, $T$, can be done numerically, providing a solution of the time-dependent Bloch vector for a given pulse sequence. By induction, it is straightforward to extend the above analysis to the general case with $n_T$ fluctuators, thus we have obtained all the necessary ingredients for an exact solution to the problem with any number of fluctuators. We note, however, that matrix size grows exponentially as $3\times 2^{n_T}$, and above $n_T=10$, exact diagonalization is computationally intensive. For larger $n_T$, we find that direct simulation of the multiple RTN is numerically more efficient, although it requires increasing number of random sampling as $n_T$ increases. The above limitations motivate us to seek approximate analytical solutions, which are obtained for the weak and strong coupling regimes, providing powerful tools that are particularly useful for larger ensembles of fluctuators. In the following subsections we first present single-TLF analytical solutions, subsequently building on them to derive the multi-TLF solutions.

\subsection{Analytical Solutions for Weak Coupling}

Here we present analytical solutions for the case of weakly-coupled TLFs, $v_i \ll \gamma_i \,\, \forall i$, accurate to second order in $v_i/\gamma_i$. For the PDD sequence we detail an explicit multi-TLF solution limited to the OP ($\theta=\pi/2$) whereas the solution for the more commonly used (and more effective) CPMG sequence is good for arbitrary qubit working point. We assume that the qubit state is prepared along the $z$ axis. Notice that this choice affects only the weights of the various decay rates in the solution, as detailed below. The perturbative solutions of the coupled set, Eqs.~(\ref{r14full}) for the two-TLF case, and of larger sets of equations, when more than two TLFs are present, rely on the single-TLF solution. In order to allow us to introduce the many-TLF solutions in a self-contained and accessible manner, we first present the solutions for the single TLF case.

\subsubsection{Single fluctuator}

The single-TLF problem was worked out in Ref.~\onlinecite{Ramon2} for a rotated reference frame whose axes are the qubit eigenstates. In the rotated frame the qubit evolves under a static field in the $z$ axis, with noise in both $x$ and $z$ axes. Here and throughout the paper, all our solutions are given in the original non-rotated frame, avoiding confusion with the components of the initial qubit state and control pulses rotation axis.

For a qubit at general working position, coupled to a single TLF, the evolution operator for a full control cycle, $T$, is given by a $6\times 6$ matrix. Performing a second-order perturbation in $v/\gamma$, we can analytically diagonalize $T$. The solution for the Bloch vector components can be generally written as:
\begin{equation}
j(t)=\sum_{i=1}^6 w_i^j e^{-\Gamma^{(i)} t}, \hspace{1 cm} j=x,y,z,
\label{jt}
\end{equation}
where the decay rates are found from the eigenvalues of $T$, $\Gamma^{(i)}= -\ln | \lambda_i| /\tau$, and $w_i^j$ is the weight of the $i$th decay rate in the solution of the $j$th component.
These weights are found from the eigenvectors of $T$, analogously to Eq.~(\ref{STS}).

For the PDD sequence, three of the six eigenvalues have nonzero weights in the solution, and the corresponding decay rates are:
\begin{eqnarray}
\Gamma^{(1)} \!&\!=\!&\! \! \frac{\gamma v^2}{\Delta^2 +4 \gamma^2} \left[ (1\!-\!A\!-\!B_1) \sin^2 \theta + C \cos^2 \theta \right] \notag \\
\Gamma^{(2,3)} \!&\!=\!&\! \! \frac{\gamma v^2}{\Delta^2 +4 \gamma^2} \left[ (2\!-\!B_1\!-\!B_2) \sin^2 \theta -F \right. \notag \\
&& \left. \pm \sqrt{F^2+D^2 \sin^2 2 \theta} \right],
\label{GPDD}
\end{eqnarray}
where the different functions of $\tau$ are given by
\begin{eqnarray}
A\!&\!=\!&\!\frac{\Delta^2-4 \gamma^2}{\Delta^2+4 \gamma^2} {\rm sinc} \tilde{\tau} \notag \\
B_1 \!&\!=\!&\! \frac{8 \gamma^2}{\Delta^2 +4\gamma^2} \cos^2 \frac{\tilde{\tau}}{2} \frac{\tanh \gamma \tau}{\gamma \tau} \notag \\
B_2 \!&\!=\!&\! \frac{8 \gamma^2}{\Delta^2 +4\gamma^2} \sin^2 \frac{\tilde{\tau}}{2} \frac{\coth \gamma \tau}{\gamma \tau} \notag \\
C \!&\!=\!&\! \frac{\Delta^2+4 \gamma^2}{2\gamma^2} \left( 1-\frac{\tanh \gamma \tau}{\gamma \tau} \right) \notag \\
D \!&\!=\!&\! \cos \frac{\tilde{\tau}}{2} \frac{\tanh \gamma \tau}{\gamma \tau} -{\rm sinc} \frac{\tilde{\tau}}{2} \notag \\
F\!&\!=\!&\! \frac{1}{2} \left[ (1-A-B_1)\sin^2 \theta -C \cos^2 \theta \right],
\label{AF}
\end{eqnarray}
and ${\rm sinc}\tilde{\tau}\equiv\sin \tilde{\tau}/\tilde{\tau}$, $\tilde{\tau} \equiv \Delta \tau$ being the normalized time interval between pulses. With the qubit initially prepared along the $z$ axis, the weights of the three rates in the PDD solution for the longitudinal ($z$) and transverse ($y$, $x$) components read:
\begin{eqnarray}
w_1^z \!&\!=\!&\! \sin^2 \theta \sin^2 \frac{\tilde{\tau}}{2} \notag \\
w_{2,3}^z \!&\!=\!&\! \frac{1}{2} \left[ \sin^2 \theta \cos^2 \frac{\tilde{\tau}}{2} \left(1 \mp \frac{F}{\sqrt{F^2+D^2 \sin^2 2 \theta}} \right) +\right. \notag \\
&& \left. \cos^2 \theta \left(1 \pm \frac{F}{\sqrt{F^2+D^2 \sin^2 2 \theta}} \right) \pm \right. \notag \\
&& \left. \sin^2 2\theta \cos \frac{\tilde{\tau}}{2} \frac{D}{\sqrt{F^2+D^2 \sin^2 2\theta}} \right] \label{WPDD}
\end{eqnarray}
\begin{eqnarray}
w_1^y \!&\!=\!&\! -\frac{1}{2} \sin \theta \sin \tilde{\tau} \notag \\
w_{2,3}^y \!&\!=\!&\! \frac{1}{4} \sin \tilde{\tau} \left( 1\mp \frac{F}{\sqrt{F^2+D^2 \sin^2 2\theta}} \right) \sin \theta \pm \notag \\
&& \frac{D \cos \theta \sin 2\theta}{2 \sqrt{F^2+D^2 \sin^2 2\theta}} \sin \frac{\tilde{\tau}}{2}
\label{WPDDy}
\end{eqnarray}
\begin{eqnarray}
w_1^x \!\!&\!=\!&\!-\frac{1}{2} \sin 2\theta \sin^2 \frac{\tilde{\tau}}{2} \notag \\
w_{2,3}^x \!\!&\!=\!&\!\frac{1}{4} \sin 2\theta \! \left[ \sin^2 \frac{\tilde{\tau}}{2} \pm \frac{F}{\sqrt{F^2\!+\! D^2 \sin^2 2\theta}} \!\left(\! 1\! +\cos^2 \frac{\tilde{\tau}}{2}\right)\! \right] \notag \\
&& \mp \frac{D\sin 4\theta}{4\sqrt{F^2+D^2 \sin^2 2\theta}}\cos\frac{\tilde{\tau}}{2}.
\end{eqnarray}
These results simplify at the OP ($\theta=\pi/2$), where we find two distinct rates:
\begin{eqnarray}
\Gamma^{(1)}(\theta=\pi/2)\!&\!=\!&\! \frac{\gamma v^2}{\Delta^2 +4\gamma^2} (1-A-B_1) \notag \\
\Gamma^{(2)}(\theta =\pi/2)\!&\!=\!&\! \frac{\gamma v^2}{\Delta^2 +4 \gamma^2} (1+A-B_2),
\label{GPDDop}
\end{eqnarray}
and the relevant Bloch vector components read:
\begin{eqnarray}
z_{PDD}^{OP}(t)\!&\!=\!&\! \sin^2 \frac{\tilde{\tau}}{2} e^{-\Gamma^{(1)} t}+\cos^2 \frac{\tilde{\tau}}{2} e^{-\Gamma^{(2)}t} \\
y_{PDD}^{OP}(t)\!&\!=\!&\! \frac{1}{2} \sin \tilde{\tau} \left( e^{-\Gamma^{(2)} t}- e^{-\Gamma^{(1)}t}
\right).
\label{ypdd}
\end{eqnarray}

Turning to the CPMG case, only two eigenvalues have nonzero weights with corresponding decay rates:\cite{rotated}
\begin{eqnarray}
\tilde{\Gamma}^{(1)} \!&\!=\!&\! \! \frac{\gamma v^2}{\Delta^2 +4 \gamma^2} \left[ (1\!+\!A\!-\!B_2) \sin^2 \theta \right. \notag \\
&& \left. + (C +2D) \cos^2 \theta \right] \notag \\
\tilde{\Gamma}^{(2)} \!&\!=\!&\! \! \frac{\gamma v^2}{\Delta^2 +4 \gamma^2} (2\!-\!B_1\!-\!B_2+2D) \sin^2 \theta,
\label{GCP}
\end{eqnarray}
where the different functions are given by Eqs.~(\ref{AF}). The associated weights in the longitudinal component read:
\begin{eqnarray}
w_1^z\!&\!=\!&\! \sin^2 \theta \notag \\
w_2^z\!&\!=\!&\! \cos^2 \theta.
\label{WCP}
\end{eqnarray}
At the OP, there is a single decay rate, $\tilde{\Gamma}^{(1)}=\Gamma^{(2)}$. Since $\Gamma^{(1)}$ is always larger than $\Gamma^{(2)}$, these results reaffirm the superior performance of the CPMG protocol. This improvement is more pronounced for slow TLFs, $\gamma \ll \Delta$, for which $\Gamma^{(1)} \gg \Gamma^{(2)}$. As one moves away from the OP, the $C$ term becomes the dominant contribution in the decay rates, a tendency that is more pronounced for slow TLFs. Since this term is present in both PDD and CPMG decay rates, the CPMG advantage is largely lost outside the OP.

The dynamics of the transverse components are typically less pronounced for CPMG as compared with PDD, since leading terms in $v/\gamma$ are canceled. Performing a calculation analogous to the one outlined in Eq.~(\ref{CP}), we find that one cannot neglect the small contributions from the $2\gamma$ decay rates,\cite{delta} as was done for the $z$ components, requiring us to work with the full $6 \times 6$ matrices. Here we focus on the OP, for which a qubit initially prepared along the $z$ axis has no $x$ dynamics, and dissipative dynamics are manifested only through the $y$ component, allowing us to consider smaller $3 \times 3$ matrices. The relevant decay rates are thus $\Gamma^{(1)}, \Gamma^{(2)}$, given in Eqs.~(\ref{GPDDop}), and $2\gamma$, and the $y(t)$ component is found by:
\begin{equation}
\left( \begin{array}{c} y \\ z \\ \delta x \end{array} \right)= \tilde{L}^{N-1} T_{1/2} T^{N-1} T_{1/2}
\left( \! \begin{array}{c} 0 \\ 1 \\ 0 \end{array} \! \right),
\label{CP3}
\end{equation}
where $T$ and $T_{1/2}$ are the evolution operators during $\tau$ and $\tau/2$ pulse intervals, respectively (see section II.A), and $\tilde{L}={\rm diag} (\pm 1,1,1)$, with the upper (lower) sign corresponding to odd (even) number of pulses. We note that while $z(t)$ is unaffected by $\tilde{L}$, even-odd effects do appear in the transverse component, as demonstrated below. Keeping terms to second order in $v/\gamma$, we find
\begin{eqnarray}
y_{CP}^{OP}(t) \!&\!=\!&\! \tilde{w}_y e^{-\Gamma^{(1)} t}-\frac{\tau}{4} \sin \frac{\tilde{\tau}}{2} \left( \Gamma_{1/2}^{(2)}-\Gamma_{1/2}^{(1)} \right) \times \notag \\
&& \left[(-1)^{N+1}e^{-\Gamma^{(2)} t}+e^{-\Gamma^{(1)} t}\right],
\label{ycp}
\end{eqnarray}
where
\begin{equation}
\tilde{w}_y \!=\! \frac{2\gamma v^2}{(\Delta^2\!+4\gamma^2)^2} \!\!\left(\! 1\!-\! \frac{\cos\tilde{\tau}/2}{\cosh \gamma \tau}\right)\!\! \left( \!\Delta\!-\!2\gamma \frac{\sin \tilde{\tau}/2}{\sinh \gamma \tau} \right)\!.
\label{wy}
\end{equation}
In Eq.~(\ref{ycp}), $\Gamma_{1/2}^{(j)}$ refer to decay rates, Eqs.~(\ref{GPDDop}), evaluated with $\tau \rightarrow \tau/2$. For an even number of pulses, the second term in Eq.~(\ref{ycp}) is negligible, leading to weaker qubit dissipation and improved performance, as compared with the case of odd number of pulses.

\subsubsection{Two or more fluctuators}

Considering first the two-TLF case at the OP, we apply the same perturbative approach in diagonalizing the evolution operator $T$, with $M_1$ given by Eq.~(\ref{M1OP}). Out of the six eigenvalues obtained for $T$, we find, like in the single-TLF case, that for PDD only two have nonzero weights in the solution. Initializing the qubit state along the $z$ axis, the qubit signal decay along the longitudinal axis reads:
\[
z_{PDD}^{OP}(t)=\sin^2\frac{\tilde{\tau}}{2} e^{-(\Gamma_1^{(1)} +\Gamma_2^{(1)})t}+ \cos^2\frac{\tilde{\tau}}{2} e^{-(\Gamma_1^{(2)} +\Gamma_2^{(2)})t}.
\]
Here, $\tilde{\tau}\equiv \Delta \tau$, and the decay rates, $\Gamma_i^{(j)}$, are the single-TLF rates given by Eqs.~(\ref{GPDDop}), where the subscript $i$ denotes quantities evaluated with the $i$th TLF parameters, $\gamma_i, v_i$.

At the OP, the two-TLF solution can be extended by induction to the general $n_T$ case, and it can be shown that the two-rate structure is retained, with all other eigenvalues having no weight in the final solution. The weights of the remaining two rates are independent of the TLF parameters, allowing us to write the general $n_T$ solution as:\cite{3TLF}
\begin{eqnarray}
z_{PDD}^{OP}(t)\!&\!=\!&\! \sin^2\frac{\tilde{\tau}}{2} e^{-\Gamma^{(1)}t}+ \cos^2\frac{\tilde{\tau}}{2} e^{-\Gamma^{(2)}t} \label{PDDn} \\
\Gamma^{(j)}\!&\!\equiv\!&\! \sum_{i=1}^{n_T} \Gamma_i^{(j)}, \hspace{0.5 cm} j=1,2. \notag
\end{eqnarray}
Outside the optimal point, there are three single-TLF rates given in Eqs.~(\ref{GPDD}), and their TLF-dependent weights for PDD, Eqs.~(\ref{WPDD}), do not allow us to simply group them in the multi-TLF case, as was done in Eq.~(\ref{PDDn}). Indeed, examining the weights of the various eigenvalues in the multi-TLF PDD solution outside the optimal point, we find that the number of contributing terms grows exponentially with $n_T$, and the general analytical solution is intractable.

In contrast, the single-TLF CPMG solution, Eqs.~(\ref{GCP}) and (\ref{WCP}), includes only two rates with weights that are independent of TLF parameters, allowing us to extend the solution to the $n_T$ case at an arbitrary working point:
\begin{eqnarray}
z_{CP}(t)\!&\!=\!&\! \sin^2\theta e^{-\tilde{\Gamma}^{(1)}t}+ \cos^2\theta e^{-\tilde{\Gamma}^{(2)}t} \label{CPn} \\
\tilde{\Gamma}^{(j)}\!&\!\equiv \!&\! \sum_{i=1}^{n_T} \tilde{\Gamma}_i^{(j)}, \hspace{0.5 cm} j=1,2, \notag
\end{eqnarray}
where we use tilde to denote the CPMG rates, which are different from the PDD rates in the general working point [compare Eqs.~(\ref{GPDD}) with Eqs.~(\ref{GCP})]. At the OP, this solution reduces to a single decay rate, $\tilde{\Gamma}^{(1)}=\Gamma^{(2)}$. Inspecting these solutions, we observe that they are, in general, not factorizable to the single-TLF solutions, as was the case at pure dephasing, Eq.~(\ref{chint}). The extent to which the exact solution deviates from the factorized single-TLF solutions, $\prod_{i=1}^{n_T} z_i (t)$, indicates the role of collective effects within the fluctuator bath in the qubit dynamics. In other words, it allows us to quantify to what extent does qubit dephasing due to one fluctuator depend on the presence of other fluctuators.

Following Ref.~\onlinecite{BerglinT}, we explain these collective effects by recalling the nonlinear dependence of the Hamiltonian eigenvalues in the qubit-TLF couplings, Eq.~(\ref{Hameigs}). At short times, $\gamma t \ll 1$, these eigenvalues can be expanded to include linear and quadratic coupling terms, each resulting in a contribution to the qubit dephasing.\cite{Cywinski2,Schriefl}
As one approaches the OP, the linear coupling contribution, which is proportional to $\cos \theta$ [see Eq.~(\ref{Hameigs})] becomes smaller, making the contribution of the quadratic coupling dominant. The physical explanation to the nonlinear contribution is that the OP for one fluctuator is no longer well defined in the presence of other TLFs, thus although the TLFs are independent RTN sources, their contributions to the qubit dephasing are not. Outside pure dephasing, slow fluctuators can thus play an important role, by introducing quasi-static changes to the OP location, thereby enhancing the effects of other fluctuators. At longer times, qubit dephasing can no longer, in general, be split into linear and quadratic contributions, and one needs to evaluate it from a single nonlinear coupling term, but the qualitative picture given above holds true, as demonstrated by the results below.

\begin{figure}[tb]
\epsfxsize=1\columnwidth
\vspace*{-0.0 cm}
\centerline{\epsffile{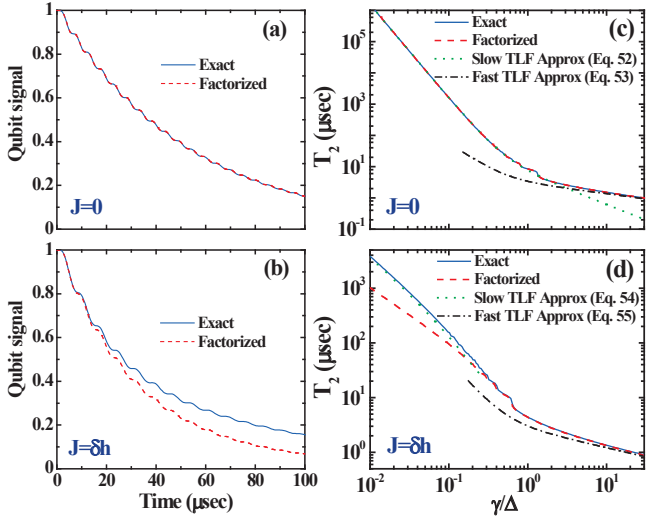}}
\vspace*{-0.1 cm}
\caption{(color online) Qubit dephasing due to four identical weakly coupled TLFs under 10-pulse CPMG. Both exact (solid-blue lines) and product of single-TLF solutions (dashed-red lines) are shown. Figures (a) and (b) capture the longitudinal ($z$) signal decay at the optimal point ($J=0$) and at $\theta =\pi/4$ ($J=\delta h=0.1 \mu$eV), respectively, with $\gamma_i=0.2 \mu {\rm sec}^{-1}$ and $v_i=2$ neV. Figures (c) and (d) depict the corresponding $T_2$ time dependence on the ratio $\gamma/\Delta$, where $v/\gamma$ is kept constant (maintaining weak coupling) by simultaneously sweeping $\gamma$ and $v$. Also included are slow-TLF (dotted green lines) and Fast-TLF (dash-dotted black lines) approximations, discussed in the main text.}
\label{Fig4}
\end{figure}
In Fig.~\ref{Fig4}(a) we plot the longitudinal ($z$) Bloch vector component at OP, subjected to four identical weakly-coupled TLFs under 10 pulse CPMG. The figure shows the full matrix diagonalization solution (blue solid line) and an approximate factorized solution (red dashed line). As expected for CPMG at OP, where dephasing is governed by a single decay rate, the two solutions are identical. In contrast, the PDD solution (not shown), which includes two decay rates, factorizes only for slow TLFs, $\gamma_i \ll \Delta$, where the two rates are approximately identical.

It is instructive to examine the asymptotic behavior of these solutions for slow and fast TLFs. For slow TLFs, satisfying $\max(\gamma_i) \ll \Delta$, PDD and CPMG perform the same with a single decay rate, independent of the number of control pulses. The resulting dephasing time is:
\begin{equation}
T_2^{ST,OP}\!\! \approx \!\! \left(\sum_{i=1}^{n_T} \frac{\gamma_i v_i^2}{\Delta^2+4\gamma_i^2}\!\! \right)^{\!\!\!\!-1}\!\!\!\! \ln \!2  \stackrel{{\rm Id. TLFs}}{\longrightarrow} \! \frac{\Delta^2\!+4\gamma^2}{\gamma v^2 n_T} \ln \! 2,
\label{stop}
\end{equation}
where we included the simplified result for $n_T$ identical TLFs. For fast TLFs, $\min(\gamma_i) \gg \Delta$, we expand $\Gamma^{(2)}$ in Eqs.~(\ref{GPDDop}) by taking $\gamma \tau \gg 1$ and $\Delta \tau \ll 1$. The resulting dephasing time is:
\begin{eqnarray}
T_2^{FT,OP} \!&\! \approx \!&\! \left[ \frac{3 N^2 \ln 2}{2\Delta^2 \sum_i v_i^2 \gamma_i^3/(\Delta^2+4 \gamma_i^2)^2} \right]^{1/3} \stackrel{{\rm Id. TLFs}}{\longrightarrow} \notag \\
 &&\left[ \frac{3 N^2\ln 2 (\Delta^2+4 \gamma^2)^2}{2 \Delta^2v^2\gamma^3 n_T}\right]^{1/3},
\label{ftop}
\end{eqnarray}
and we recover the $N^{2/3}$ power law predicted \cite{Cywinski,Ramon2} and observed\cite{Lange,Medford} in previous works. The dependence of $T_2$ times on the TLF switching rate $\gamma$ (identical for all four TLFs) is depicted in Fig.~\ref{Fig4}(c), where the TLF coupling strength $v$ is swept along with $\gamma$ to maintain a constant $v/\gamma$ ratio within the weak coupling regime. The figure shows the asymptotes, Eqs.~(\ref{stop}) and (\ref{ftop}), and the factorized single-TLF solution, identical to the exact solution for CPMG at OP.

Outside the OP, the CPMG solution includes two comparable but non-identical decay rates, thus it is no longer factorizable and collective effects begin to show up, as demonstrated in Fig.~\ref{Fig4}(b) for $J=\delta h$ ($\theta=\pi/4$). Examining the asymptotic behavior at the general working point for the slow-TLF case, $\max(\gamma_i) \ll \Delta$, we have two distinct single-TLF decay rates: $\Gamma^{(1)}_i \approx \cos^2 \theta v_i^2/2 \gamma_i$ and $\Gamma^{(2)}_i \approx 2\sin^2 \theta \gamma_i v_i^2/\Delta^2$. The dephasing time can be generally found as:
\begin{eqnarray*}
T_2^{ST}\!&\!=\!&\!\frac{1}{\Gamma^{(2)}} \left[ 1-\frac{1}{2} \sec^2 \theta+ \right.\\
&&\!\! \left. \frac{\Gamma^{(2)}}{\Gamma^{(1)}} W \!\left( \frac{\Gamma^{(1)}}{\Gamma^{(2)}} e^{-\frac{\Gamma^{(1)}}{\Gamma^{(2)}}(1-\frac{1}{2} \sec^2 \theta) } \tan^2 \theta  \right) \right],
\end{eqnarray*}
where $W(z)$ is the Lambert W function, solving the equation $z=W(z) e^{W(z)}$. As long as we are not too close to the OP, $\Gamma_i^{(1)} \gg \Gamma_i^{(2)}$ $\forall i$, and for $\theta=\pi/4$, the dephasing time is found to first order in $\Gamma^{(2)}/\Gamma^{(1)}$ as:
\begin{eqnarray}
T_2^{ST}\!&\!=\!&\! \frac{1}{\Gamma^{(1)}} \ln \left[ \frac{\Gamma^{(1)}/\Gamma^{(2)}}{\ln (\Gamma^{(1)}/\Gamma^{(2)})} \right] \stackrel{{\rm Id. TLFs}}{\longrightarrow}  \notag \\
&& \frac{4\gamma }{n_T v^2} \ln \left[ \frac{\Delta^2 }{8\gamma^2 \ln(\Delta /2\gamma)} \right].
\label{Slow}
\end{eqnarray}
Notice that in this limit, the initial decay is governed by the fast $\Gamma^{(1)}$ until the signal has dropped to $\approx 50\%$ [See Eq.~(\ref{CPn}) with $\theta =\pi/4$], after which, a much slower decay, $\Gamma^{(2)}$, takes place. In this regime, exhibiting two very different decay rates, collective effects are strongest, as indicated by the large deviation of the factorized solution from the exact solution in Fig.~\ref{Fig4}(d). In many practical situations, one is interested in coherence times for which the signal remains above, say, $95\%$ of its initial value. In this slow TLF regime, the initial decay time is governed by $\Gamma^{(1)}$ thus, it is much shorter than $T_2^{ST}$ given in Eq.~(\ref{Slow}). As in the OP case, the qubit dephasing due to slow TLFs is independent of the DD protocol or the number of control pulses, as long as the time interval between pulses satisfies $\tau \ll \gamma_i^{-1}$.

For fast TLFs, we find a single effective rate, $\tilde{\Gamma}=(\tilde{\Gamma}^{(1)}+\tilde{\Gamma}^{(2)})/2$, which can be expanded as in the OP case to give:
\begin{eqnarray}
T_2^{FT} \!&\! \approx \!&\! \left[ \frac{2 \ln 2 N^2}{2\Delta^2 (1-\frac{2}{3}\cos^2\theta) \sum_i v_i^2 \gamma_i^3/(\Delta^2+4 \gamma_i^2)^2} \right]^{1/3} \notag \\
 && \stackrel{{\rm Id. TLFs}}{\longrightarrow}  \left[ \frac{2\ln 2 (\Delta^2+4 \gamma^2)^2 N^2}{(1-\frac{2}{3}\cos^2\theta) \Delta^2v^2\gamma^3 n_T}\right]^{1/3}.
\end{eqnarray}
At this limit, the factorized solution coincides with the exact result, as demonstrated by the righthand side of Fig.~\ref{Fig4}(d).

We now turn our attention to the qubit dissipative dynamics, by considering the transverse components of the Bloch vector. At the OP, a qubit initially prepared along the $z$ axis has no $x$ dynamics and dissipative dynamics are manifested only through the $y$ component. Extending the single-TLF analysis presented in the previous subsection to the multi-TLF case, we find that the formulas for $y(t)$ for the PDD [Eq.~(\ref{ypdd})] and CPMG [Eq.~(\ref{ycp})] cases hold true by substituting $\Gamma^{(j)}$ and $\Gamma_{1/2}^{(j)}$ with summations over single-TLF decay rates, as was done in Eq.~(\ref{PDDn}). Similarly, $\tilde{w}_y$ in Eq.~(\ref{wy}) is replaced with summation over all TLF weights, $\sum_{i=1}^{n_T}\tilde{w}^i_y$, each evaluated with its respective TLF parameters $\gamma_i, v_i$. We note that these dissipative effects are of higher order and are thus never factorizable.

\begin{figure}[tb]
\epsfxsize=0.9\columnwidth
\vspace*{-0.0 cm}
\centerline{\epsffile{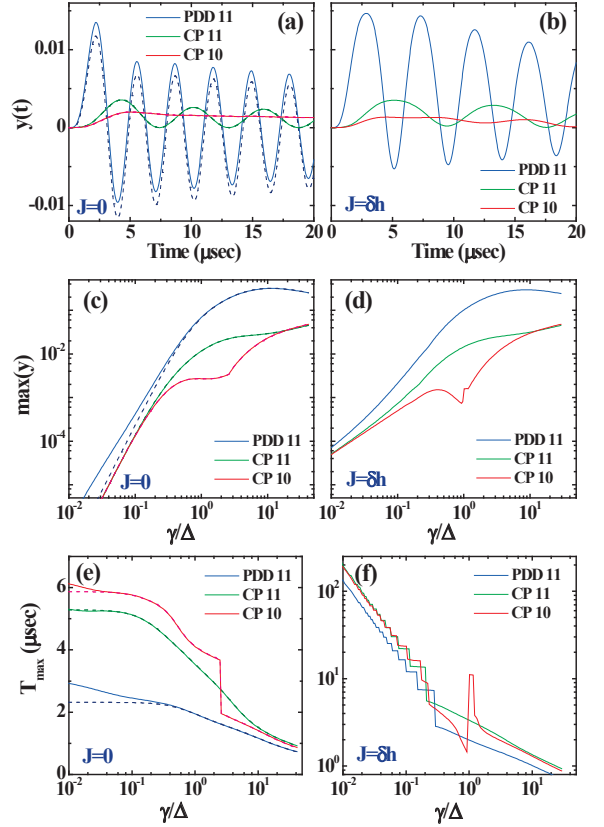}}
\vspace*{-0.1 cm}
\caption{(color online) $y(t)$ Bloch vector component of a qubit, initially prepared along the $z$ axis, due to four identical weakly coupled TLFs. Figures (a) and (b) show $y(t)$ for 11-pulse PDD, 11-pulse CPMG and 10-pulse CPMG, at the OP and at $J=\delta h=0.1 \mu$eV, respectively, with $\gamma_i=0.2 \mu {\rm sec}^{-1}$ and $v_i=2$ neV. Figures (c) and (d) depict the maximal values of $y(t)$ at the OP and at $\theta=\pi/4$ vs.~$\gamma/\Delta$, while keeping $v/\gamma$ constant. Figures (e) and (f) provide the corresponding times, at which maximal $y(t)$ values are reached. Dashed lines in Figures (a), (c) and (e) depict the analytical formulas given in Section III.A.1 C for the OP (see main text), and are compared against exact numerical diagonalization (solid lines).}
\label{Fig5}
\end{figure}

The analytical solutions for $y(t)$ at the OP are compared with the results of numerical diagonalization in Figure \ref{Fig5}(a), for four identical weakly-coupled TLFs, under 11 pulse PDD (blue line), 11 pulse CPMG (green line), and 10 pulse CPMG (red line). The accuracy of the analytical solutions for the CPMG sequences is obtained by including the $2 \gamma$ rate contribution, whereas a less accurate result is shown for the PDD case for which this contribution was neglected. In Figs.~\ref{Fig5}(c) and (e) we depict the maximum value of $y(t)$ and the time to reach that maximum, respectively, as we simultaneously vary $\gamma$ and $v$. The superior performance of the CPMG sequence is evidenced throughout the parameter range. CPMG sequences with even number of pulses are better than those with odd number of pulses [see Eq.~(\ref{ycp})], but this advantage is washed out for very slow or very fast TLFs [see Fig.~\ref{Fig5}(c)]. For slow TLFs, $v \ll \gamma \ll \Delta$, in particular, $\tau \sim 1/\Delta$, and inspection of Eqs.~(\ref{ycp}) and (\ref{wy}) leads to:
\begin{equation}
y_{\rm max}^{ST} \approx \sum_{i=1}^{n_T} \tilde{w}_y^i \approx \sum_{i=1}^{n_T} \frac{4 \gamma_i v_i^2}{\Delta^3},
\end{equation}
irrespective of the number of control pulses. This result agrees well with the slow-TLF asymptote in Fig.~\ref{Fig5}(c). Figures \ref{Fig5}(b), (d) and (f) depict the $y(t)$ dynamics at $\theta=\pi/4$ ($J=\delta h$). The most striking difference with respect to the optimal point takes place at the slow-TLF regime, where substantially larger $y$ values are obtained.

\begin{figure}[tb]
\epsfxsize=0.9\columnwidth
\vspace*{-0.0 cm}
\centerline{\epsffile{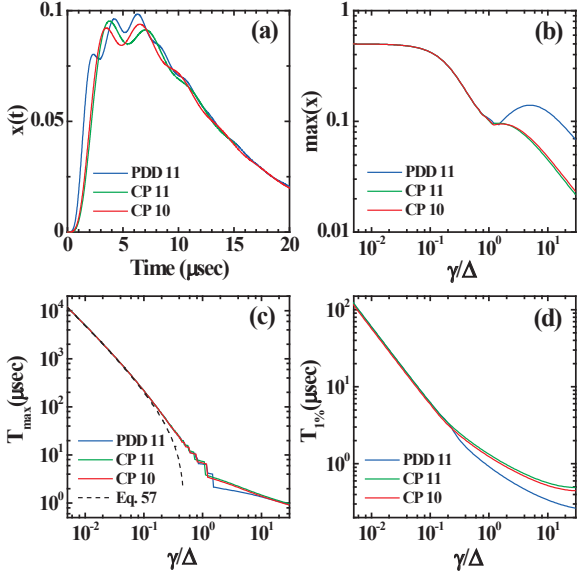}}
\vspace*{-0.2 cm}
\caption{(color online) $x(t)$ Bloch vector component of a qubit initially prepared along the $z$ axis, due to four identical weakly coupled TLFs, for $J=\delta h=0.1 \mu$eV working point. Figure (a) shows $x(t)$ for 11-pulse PDD, 11-pulse CPMG, and 10-pulse CPMG, with $\gamma_i=0.05 \mu {\rm sec}^{-1}$ and $v_i=8$ neV. Figure (b) depicts the maximal values of $x(t)$ vs.~$\gamma/\Delta$, while keeping $v/\gamma$ constant. Figure (c) shows the corresponding times, at which maximal $x(t)$ values are reached. Dashed line corresponds to the analytical result, Eq.~(\ref{Slowx}), applicable for the slow-TLF regime. Figure (d) shows times to reach $x=0.01$ vs. $\gamma/\Delta$.}
\label{Fig6}
\end{figure}

Finally, dissipative dynamics along the $x$ axis occur for a qubit initially prepared along the $z$ axis only when operating away from the optimal point. As seen in Fig.~\ref{Fig6} for working position $J=\delta h$ ($\theta=\pi/4$), a substantial buildup of $x$ component, up to 50\%, is obtained for slow TLFs, albeit at increasingly longer time scales. At this limit, where $v \ll \gamma \ll \Delta$, the dynamics are indifferent to the pulse sequence, as well as to the number of pulses. Similarly to the discussion above Eq.~(\ref{Slow}), there are two distinct single-TLF decay rates, where $\Gamma_i^{(1)} \gg \Gamma_i^{(2)}$, and the $x$ component reads $x(t)=\frac{1}{2} [e^{-\Gamma^{(2)}t}-e^{-\Gamma^{(1)}t}]$. At this slow-TLF limit we find the time to reach the maximum is:
\begin{eqnarray}
T_{{\rm max}(x)}^{ST}\!&\!=\!&\! \frac{1}{\Gamma^{(1)}} \ln \left[\Gamma^{(1)}\!/\Gamma^{(2)} \right] \stackrel{{\rm Id. TLFs}}{\longrightarrow}  \notag \\
&& \frac{4\gamma \sec^2 \theta}{n_T v^2} \ln \left( \frac{\Delta \cot \theta}{2\gamma} \right).
\label{Slowx}
\end{eqnarray}
Eq.~(\ref{Slowx}) is depicted by the dashed line in Fig.~\ref{Fig6}(c), and agrees well with the results of exact numerical diagonalization at the slow-TLF regime. Improved performance of the CPMG sequences over PDD is found only when $\gamma \gtrsim \Delta$ [see righthand side of Figure \ref{Fig6}(b)]. In contrast with $y(t)$ dynamics, there is no improvement in performance gained by employing an even number of CPMG pulses.

\subsection{Strong Coupling}

Similarly to the weak coupling regime, we use a perturbative approach to diagonalize the evolution operator, $T$, in the strong coupling regime, where $\gamma_i \ll v_i \,\, \forall i$. The solutions to the many-TLF problem are not directly derivable from the single-TLF solutions, as was the case for weak coupling. Below we provide results for both single- and many-TLF cases, allowing us to draw conclusions on the role of collective effects in qubit dephasing, and scaling of the noise with the number of TLFs, in the strong coupling regime.

\subsubsection{Single Fluctuator Near or at the optimal point}

For a strongly coupled TLF, it is sufficient to perform first order degenerate perturbation theory in $\gamma/v$. The analytical expressions are lengthy and we present here results for the optimal point (see Ref.~\onlinecite{Ramon2} for results for general working point in the rotated frame), expanded to third order in $v/\delta h$. At the OP, the equations decouple and we only need to solve for $(y,z,\delta x)$, similarly to the weak coupling case. The three eigenvalues of the resulting evolution operator lead to two decay rates, relevant for both PDD and CPMG protocols:
\begin{eqnarray}
\Gamma^{(1)}\!&\!=\!&\!\frac{\gamma v^2}{\Delta^2} \left( 1 -{\rm sinc} \tilde{\tau} \right) \notag \\
\Gamma^{(2)}\!&\!=\!&\!\frac{\gamma v^2}{\Delta^2} \left( 1-2{\rm sinc}^2 \frac{\tilde{\tau}}{2} +{\rm sinc} \tilde{\tau} \right),
\label{GS}
\end{eqnarray}
and a much faster third rate, $\Gamma^{(3)} \approx 2\gamma$, predominantly associated with $\delta x$. Whereas in the weak coupling regime, this latter decay rate had no effect on the qubit signal, in the strong coupling regime, its weight in the final solution is not negligible.

Taking the qubit initial state to lie along the $z$ axis, we find the longitudinal ($z$) and transverse ($y$) Bloch vector components by using Eq.~(\ref{jt}) (there is no $x$ dynamics at the OP). The weights of the three decay rates in the PDD solution are found as:
\begin{eqnarray}
w_1^z \!&\!=\!&\! \sin \frac{\tilde{\tau}}{2} \left( 1-\frac{v^2}{\Delta^2} \right) +\left( \frac{v}{2 \Delta}\right)^2  \tilde{\tau} \sin \tilde{\tau} \notag \\
w_2^z \!&\!=\!&\! \cos^2 \frac{\tilde{\tau}}{2} +\frac{v^2}{\Delta^2} \left[ \sin^2 \! \frac{\tilde{\tau}}{2}-(1-{\rm sinc} \tilde{\tau})^2 \!-\!\frac{\tilde{\tau}}{4}\! \sin \tilde{\tau} \right] \notag \\
w_3^z \!&\!=\!&\! \frac{v^2}{\Delta^2} \left( 1-{\rm sinc} \tilde{\tau} \right)^2
\label{wzs}
\end{eqnarray}
\begin{eqnarray}
w_1^y \!&\!=\!&\! -\frac{1}{2} \sin \tilde{\tau} -\left(\frac{v}{2\Delta}\right)^2 \tilde{\tau} \left( \cos \tilde{\tau} - {\rm sinc} \tilde{\tau} \right) \notag \\
w_2^y \!&\!=\!&\! \frac{1}{2} \sin \tilde{\tau} \!+\! \left( \frac{v}{2\Delta}\right)^2 \!\! \left[ \! 2 {\rm sinc}^2 \frac{\tilde{\tau}}{2} (\tilde{\tau}\!-\! \sin \tilde{\tau}) \!+ \tilde{\tau} \! \cos \tilde{\tau}\!-\!\sin \tilde{\tau} \! \right] \notag \\
w_3^y \!&\!=\!&\! -\frac{v^2}{2\Delta^2} \tilde{\tau} {\rm sinc}^2\frac{\tilde{\tau}}{2} \left( 1-{\rm sinc} \tilde{\tau} \right).
\label{wys}
\end{eqnarray}

For CPMG, we find, similarly to the weak-coupling case (see Section III.A.1), that the faster decay rate, $\Gamma^{(1)}$, is eliminated from the longitudinal component. The improved performance, as compared with PDD, is nevertheless compromised in the strong coupling, due to the contribution of the fast decay rate $\Gamma^{(3)}=2\gamma$ that is also present in the CPMG solution. The weights of these two remaining rates are found as:
\begin{eqnarray}
w_2^z \!&\!=\!&\! 1-\frac{v^2}{\Delta^2} \left(1-{\rm sinc} \frac{\tilde{\tau}}{2} \right)^2 \notag \\
w_3^z \!&\!=\!&\! \left(1-{\rm sinc} \frac{\tilde{\tau}}{2} \right)^2,
\end{eqnarray}
and the solution for the transverse $y$ component is found to be:
\begin{eqnarray}
y_{CP}^{OP} \!&\!=\!&\! \frac{4\gamma v^2}{\Delta^3} \sin^2 \frac{\tilde{\tau}}{4} \left[ \sin^2 \frac{\tilde{\tau}}{4} \left(e^{-\Gamma^{(1)} t} -(-1)^N e^{-\Gamma^{(2)} t} \right) + \notag \right. \\
&& \left. (-1)^N \left(1-{\rm sinc} \frac{\tilde{\tau}}{2} \right) \left(e^{-\Gamma^{(2)} t} -e^{-2\gamma t} \right) \right]
\label{ycps}
\end{eqnarray}

\begin{figure}[t]
\epsfxsize=0.95\columnwidth
\vspace*{0.0 cm}
\centerline{\epsffile{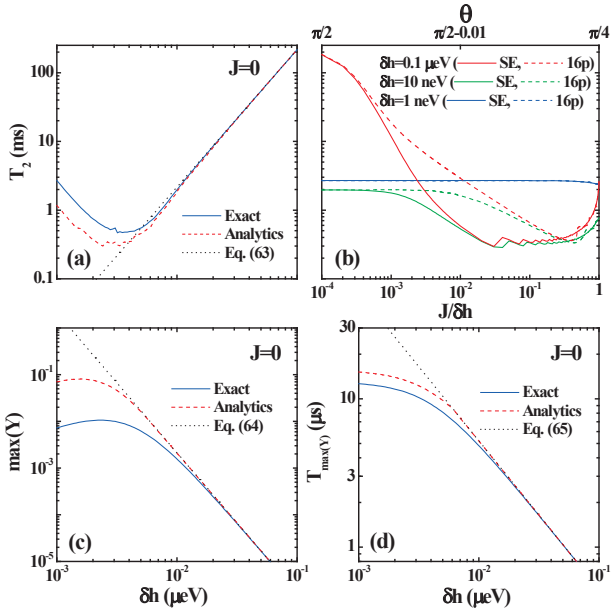}}
\vspace*{-0.2 cm}
\caption{(color online) Qubit dephasing due to a strongly coupled TLF ($v=2$ neV, $\gamma^{-1}=0.5$ msec). Figure (a) shows $T_2$ time vs.~$\delta h$ for SE, at the optimal point ($J=0$). Figure (b) shows $T_2$ times as one moves away from the optimal point by ramping up $J$ to $\delta h$ ($\theta=\pi/4$, see top axis). Three different $\delta h$ values are depicted for both SE and 16 pulse CPMG protocols.  Figures (c) and (d) capture the maximum value of the transverse ($y$) component and the corresponding time, respectively, vs.~$\delta h$ for SE, at the optimal point. The dashed red lines in figures (a), (c), and (d) depict the full analytical results of Eqs.~(\ref{GS}), (\ref{wzs}), and (\ref{ycps}), whereas the dotted black lines show the single-rate approximations.}
\label{Fig7}
\end{figure}

Figure \ref{Fig7}(a) depicts SE dephasing time vs.~$\delta h$ due to a single strongly-coupled TLF, at the optimal point. For $\delta h \ll v \ll \gamma$, the magnetic field gradient provides protection against the noise, and very long coherence times, in excess of 100 msec, are obtained. At this limit, $\tau \ll \delta h^{-1}$ and  Eqs.~(\ref{GS}) reduce to a single decay rate: $\Gamma \approx 4 \gamma v^2/\delta h^2$. This single-exponential decay, results in approximate dephasing time:
\begin{equation}
T_2 \approx \frac{\ln 2}{4\gamma} \left(\frac{\delta h}{v}\right)^2,
\end{equation}
shown by the black dotted line, and is independent of the number of control pulses. This approximation holds well down to $\delta h \sim 10v$, while the full analytical results given by Eqs.~(\ref{GS}) and (\ref{wzs}) are valid for an extended regime (strictly they are correct to third order in $v/\delta h$). As $\delta h$ becomes comparable to $v$, dephasing times drop by more than two orders of magnitude, remaining indifferent of the number of control pulses up to $N=100$. We note that nuclear polarization cycles have been successfully employed to generate stabilized interdot field gradients in excess of 5 mT,\cite{BluhmPRL, Shulman} well above $\delta h=0.1 \mu eV$, demonstrated here to induce extended coherence at the optimal point.

Figure \ref{Fig7}(b) examines the robustness of the noise immunity given by $\delta h$, as one moves away from the optimal point. For the $S-T_0$ qubit, this is particularly relevant, since a convenient idle point in this system is at large negative detuning, where $J$ is as small as a few neV but not strictly zero. For relatively large $\delta h=0.1 \mu$eV (red lines), the long coherence times of over 200 msec, obtained at the optimal point, drop rapidly by a factor of 500 with $J=1$ neV, only 1\% of $\delta h$. Dephasing times retain their order of magnitude thereafter all the way to $J=\delta h$ ($\theta =\pi/4$). Increasing the number of control pulses extends the regime of enhanced coherence (dashed red line). As $\delta h$ reduces, the noise immunity is gradually removed, as well as the sensitivity to the qubit working position. For $\delta h \lesssim v$ (blue lines), dephasing times are virtually indifferent to change in $J$.

Figures \ref{Fig7}(c) and (d) depict the maximum value of the transverse ($y$) Bloch vector component and the time to reach it, respectively, vs.~$\delta h$. At the limit $\delta h \ll v \ll \gamma$, where $\Gamma^{(1)} \approx \Gamma^{(2)}$, Eq.~(\ref{ycps}) takes a simple form and we find
\begin{eqnarray}
y_{\rm max} \!&\!=\!&\! \frac{16\gamma v^2}{\delta h^3} \left(3-(-1)^N \right) \label{ymaxs} \\
T_{{\rm max}(y)} \!&\!=\!&\! \frac{4\pi N}{\delta h}. \label{tmaxs}
\end{eqnarray}
This result suggests a factor of 2 reduction in the transverse component amplitude with an even number of CPMG control pulses, as compared  with odd $N$, similarly to the even-odd effect found for the weak-coupling regime (see section III.A.1). We note that, while Eq.~(\ref{ymaxs}) works well for both even and odd extended pulse sequences, Eq.~(\ref{tmaxs}) is strictly correct only for SE. The competition between the two terms in Eq.~(\ref{ycps}) results in oscillatory behavior that typically has non-monotonous amplitude for extended pulse protocols. Eq.~(\ref{tmaxs}) reflects the position of the first maximum, which is also the global maximum for the SE case, but not necessarily so for longer $N$-pulse sequences.

\subsubsection{Two or more fluctuators}

In the strong coupling regime the sensitivity of the contribution of one TLF to qubit dephasing to switchings of other TLFs is heightened, resulting in striking collective effects. The scaling of noise with the number of TLFs is, therefore, nontrivial and the individual TLF decay rates do not simply add up as in the weak coupling case. Here we present both numerical and analytical results for identical TLFs, restricting our analytical results to the optimal point, and to the quasi-static regime, $\delta h\gg v \gg \gamma$, where single-TLF formulas were provided in the previous subsection.

Considering first the two-TLF case, we follow the procedure presented above, by diagonalizing the evolution matrix, $T=Le^{M_1 \tau}$, where $M_1$ is given in Eq.~(\ref{M1OP}). We employ first order perturbation theory in $\gamma/v$, and expand our results to third order in $v/\delta h$. The resulting PDD solution for two identical TLFs can be approximated with two decay rates:
\begin{eqnarray}
\Gamma^{(1)} \!&\!=\!&\! 2\gamma \left[ 1+{\rm sinc} \left( \frac{v^2\tau}{\Delta}\right) \right] \notag \\
\Gamma^{(2)} \!&\!=\!&\!  2\gamma \left[ \frac{v^2}{\Delta^2}+1-{\rm sinc} \left(\frac{v^2\tau}{\Delta} \right) \right],
\label{G2TLFs}
\end{eqnarray}
with corresponding weights:
\begin{eqnarray}
w_1^z \!&\!=\!&\! \sin^2 \left( \frac{v^2 \tau}{2\Delta}\right) \notag \\
w_2^z \!&\!=\!&\! \cos^2 \left( \frac{v^2 \tau}{2\Delta} \right).
\label{w2TLFs}
\end{eqnarray}
For CPMG, the weight of the faster decay rate, $\Gamma^{(1)}$, is eliminated and we obtain, as before, a single-rate dephasing:
\begin{equation}
z_{CP}\approx e^{-\Gamma^{(2)} t},
\label{zCPs}
\end{equation}
with superior performance as compared with PDD.

\begin{figure}[t]
\epsfxsize=0.95\columnwidth
\vspace*{-0.0 cm}
\centerline{\epsffile{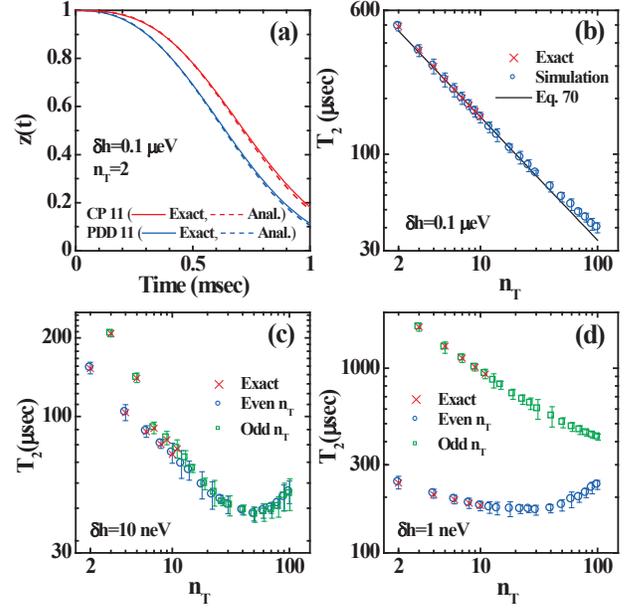}}
\vspace*{-0.2 cm}
\caption{(color online) Qubit dephasing due to two or more identical TLFs in the strong coupling regime, at the optimal point ($J=0$). (a) Bloch longitudinal component decay due to two identical TLFs, under 11-pulse PDD (Blue lines) and 11-pulse CPMG (red lines) sequences. Solid lines depict the results of exact diagonalization, whereas dashed lines correspond to Eqs.~(\ref{G2TLFs}), (\ref{w2TLFs}), and (\ref{zCPs}). Plots (b)-(d) depict scaling of $T_2$ times with the number of TLFs under 6-pulse CPMG, for three values of $\delta h$. Red crosses represent results of exact diagonalization up to $n_T=11$, and Blue circles and Green squares represent the results of multiple RTN simulations, detailed in the main text. Figure (b) also shows the analytical single-rate approximation (black line), valid for $\delta h \gg n_T v$. The TLFs parameters in all plots are: $\gamma^{-1}=0.5$ msec and $v=2$ neV.}
\label{Fig8}
\end{figure}

Figure \ref{Fig8}(a) depicts qubit dephasing at the OP, due to two identical TLFs with $\delta h \gg v \gg \gamma$, where the approximate solutions given by Eqs.~(\ref{G2TLFs})-(\ref{w2TLFs}) (dashed blue line), and Eq.~(\ref{zCPs}) (dashed red line) are compared against exact numerical diagonalization. Coherence under 11-pulse CPMG is somewhat improved ($T_2=0.72$ ms) as compared with 11-pulse PDD ($T_2=0.64$ ms). The most important observation, though, is a dramatic, 300-fold reduction in dephasing time, as compared with the single TLF case, shown in Figure \ref{Fig7}(a) (recall that at the OP, in the limit $\delta h \gg v$, single-TLF dephasing is indifferent to the number of control pulses). Evidently, the addition of a second (and subsequent) TLF(s) results in a loss of the noise immunity provided by $\delta h$ at the OP, when it is much greater than $v$. The explanation is similar to the one given in subsection III.A.2. When two or more TLFs are present, there cannot be a single OP around which all TLFs work. Switching one TLF, effectively moves the working position away from the OP for the other TLFs, resulting in a dramatic drop in coherence time, when sensitivity to the working point is high. Such sensitivity is demonstrated for the considered case in Figure \ref{Fig7}(b), where a slight deviation from the OP in the single-TLF case results in a similarly dramatic drop in coherence time.

The generalization of the above results to $n_T>2$ is far from trivial. We have diagonalized the evolution operator for the case of three identical TLFs, using first-order degenerate perturbation, but the expressions are lengthy and not very illuminating. Analytical diagonalization beyond $n_T=3$ becomes intractable, and we find through exact numerical diagonalization that the solution comprises an increasing number of distinct decay rates with non-vanishing weight as $n_T$  grows.\cite{StrongnT} In the limit of $\delta h \gg n_T v \gg n_T \gamma$, the full analytical solutions for $n_T=2$ and $n_T=3$ (not shown) can be approximated by a single-rate decay solution, which is found to hold well up to a large number of TLFs. For PDD we find the approximate decay rate to be:
\begin{equation}
\Gamma_{PDD}=\frac{2\gamma v^4}{3\delta h^2} \tau^2 n_T^2,
\end{equation}
and for CPMG we find $\Gamma_{CP}=\Gamma_{PDD}/2$. The resulting dephasing times are
\begin{equation}
T_2^{PDD}= \left( \frac{3 \ln2 \delta h^2 N^2}{2 \gamma v^4 n_T^2} \right)^{1/3},
\label{T2PDDs}
\end{equation}
for PDD and $T_2^{CP}=2^{1/3} T_2^{PDD}$ for CPMG. Eq.~(\ref{T2PDDs}) suggests that the same $N^{2/3}$ power law, found in the weak-coupling regime holds for two or more (identical) TLFs in the strong-coupling regime.

Figures \ref{Fig8}(b)-(d) show the scaling of qubit dephasing times with the number of (identical) TLFs under 6-pulse CPMG at the OP, with $\delta h=0.1 \mu eV$, $\delta h =10$ neV, and $\delta h =1$ neV, respectively. Our exact diagonalization method is limited to $n_T=11$ due to the exponential increase of matrices size. In order to investigate the dynamics with larger ensembles, we employ a direct numerical simulation of the TLFs random switchings, using many realizations of Poisson processes for the RTN sources. The resulting dephasing times and their error bars, shown in the figures by blue circles and green squares, are obtained from the noise realization averages and their standard deviations. We have found that $10^6$ realizations are typically sufficient to provide results with errors below $5\%$, and often fewer realizations are needed. The results of these simulations compare well with the results of exact diagonalization (red crosses) for $n_T\leq 11$, giving us confidence to rely on them for larger $n_T$, where exact results are unavailable.

Figure \ref{Fig8}(b) shows $T_2$ times with large field gradient, where the power law scaling with $n_T$, predicted by the single-rate analytical result, Eq.~(\ref{T2PDDs}), is expected to be valid. This approximation holds well up to 20-30 TLFs, at which point $n_T v \approx \Delta$. For larger $n_T$, the noise effects begin to saturate. The qubit dynamics become more complicated as the field gradient is reduced in figures \ref{Fig8}(c) and (d), exhibiting a non-monotonous dependence on $\delta h$, also found for the single-TLF case [see Fig.~\ref{Fig7}(a)], as well as a surprising non-monotonous dependence on $n_T$. Moreover, an even-odd effect with respect to the number of fluctuators develops as $\delta h$ reduces, showing markedly improved performance with odd $n_T$, when $\delta h=1$ neV.\cite{StrongnT} A qualitative explanation to the latter phenomenon is as follows. Each TLF in an odd $n_T$ ensemble sees an even number of TLFs that can average their switchings such that it will approximately operate at the optimal point. In contrast, for even $n_T$ the environment of each TLF will have a leftover TLF that will cause operation further away from the OP, thereby reducing coherence time. This effect should be more pronounced for smaller ensembles, where each TLF has a more prominent role in the total dynamics, as observed in the figures. In addition, the difference between even and odd $n_T$ ensembles will be evident only when $v$ is not much smaller than $\delta h$, so that the addition of a single TLF makes a substantial difference in the field about which the qubit precesses. We also stress that this effect is not likely to be apparent for non-identical TLF ensembles, particularly those with wide parameter distributions.

\begin{figure}[t]
\epsfxsize=0.9\columnwidth
\vspace*{0.05 cm}
\centerline{\epsffile{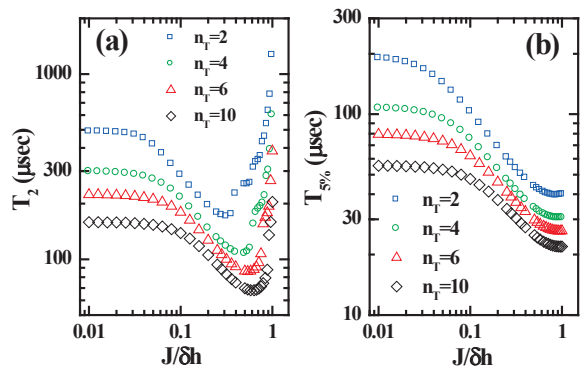}}
\vspace*{-0.2 cm}
\caption{(color online) Qubit dephasing times due to strongly-coupled identical TLFs, as a function of its working point. (a) $T_2$ times (signal drops to $50\%$; (b) Times to drop to $95\%$. Qubit Working position is varied from $J=0.01\delta h$ (very close to OP) to $J=\delta h$ ($\theta =\pi/4$). The TLFs parameters $\gamma^{-1}=0.5$ msec and $v=2$ neV, and $\delta h=0.1\mu eV$.}
\label{Fig9}
\end{figure}

Finally, in Fig.~\ref{Fig9} we examine dephasing times in the regime $\Delta \gg v \gg \gamma$, as one moves away from the OP, by ramping up J from $0.01 \delta h$ ($\theta \rightarrow \pi/2$) to $J=\delta h$ ($\theta=\pi/4$). Eq.~(\ref{T2PDDs}) works well up to $J \approx 0.05 \delta h$, at which point the scaling of noise with $n_T$ breaks down. Fig.~\ref{Fig9}(a) exhibits a non monotonous dependence on $J$, with $T_2$ times rapidly increasing as the $\theta=\pi/4$ point is approached. at this point, the qubit precesses about a tilted axis, such that its $z$ component (as well as its $x$ component) decays to 0.5 rather than to zero. This results in an artificial extension of $T_2$ times, which is accompanied by a substantial buildup of the $x$ component. This behavior is absent in the initial $5\%$ drop, shown in Fig.~\ref{Fig9}(b), as expected.

\section{Conclusion}

In this work we have studied the dynamics of a qubit coupled to a collection of two-level fluctuators, under dynamical decoupling control pulses. We have presented a theory that provides exact solutions for the Bloch vector by finding a set of coupled stochastic rate equations and diagonalizing the qubit resulting evolution operator during a full control cycle.

At pure dephasing, we obtained exact analytical solutions for asymmetric fluctuators under $N$-pulse PDD and CPMG sequences. These solutions were shown to deviate substantially from the Gaussian approximation for strongly-coupled TLFs, particularly at the long-time limit. Gaussianity was shown to be restored as the number of TLFs is increased or the inter-pulse time interval, $\tau$, becomes shorter. Lastly, we found simple formulas for noise scaling in the asymptotic limits of short time, and weak- and strong-coupling.

At the general working position, we obtained analytical solutions for the weak and strong coupling regimes, and examined collective effects within the TLF ensemble that were manifested by the deviation of the exact solution from an approximate factorized solution comprised of single-TLF solutions. These effects originate from the nonlinear nature of the qubit-TLF couplings that generates fluctuations of the OP for a given TLF due to the switchings of other TLFs. Only at pure dephasing ($\theta=0$), where the nonlinearity of the qubit-TLF couplings is eliminated, is the solution found as a product of single-TLF-induced decays. Collective effects are particularly dominant in the strong-coupling regime, due to heightened sensitivity to the qubit working point, resulting in a nontrivial noise scaling with the number of TLFs.

An additional implication of operating the qubit outside the pure dephasing regime, is the appearance of dissipative dynamics, evident by the buildup of transverse Bloch vector components. We have quantified these effects and demonstrated that the superiority of CPMG over PDD, well established at pure dephasing, extends to the general working point, with up to two-orders-of-magnitude reduction of transverse component buildup for CPMG. Furthermore, we confirmed the existence of an even-odd effect with respect to the number of CPMG control pulses, $N$, demonstrating reduced dissipation for even $N$.

In this paper, we have focused on deriving analytical solutions for the dynamics of a generic qubit system coupled to classical charge fluctuators, operating at general working position. The different asymptotic limits that we have found, as well as the decoherence scaling with ensemble size and number of control pulses, allow for a direct experimental testing, provided that the charge environment can be characterized.  Alternatively, the results of this work can aid in such an experimental characterization, as they connect measurable, qubit quantities with controllable parameters. An approach complementary to the current study would be to carry out numerical simulations of TLF ensembles with particular parameter distributions. Whereas our analysis is relevant to the mesoscopic regime with a relatively small number of TLFs, the latter approach should be useful for larger ensembles, where it is experimentally impossible to determine individual TLF parameters. It would be interesting to determine the minimal set of TLF-ensemble characteristics, necessary to predict qubit performance, or alternatively, to what extent can one characterize the charge environment, based on noise spectrum measurements.

Finally, in this paper we have used the spin-fluctuator model, treating the TLFs as classical sources of RTN. Several previous works studied the validity domain of this model in predicting qubit decoherence at pure dephasing.\cite{Abel,Wold} It would be interesting to extend these studies to the general working position by formulating a quantum telegraph model that allows for qubit energy relaxation, and to consider the effects of TLF interactions.

\section*{ACKNOWLEDGEMENTS}

This work was supported by the National Science Foundation Grant no.~DMR 1207298.

\section*{APPENDIX A: COMPARISON WITH CUMULANT EXPANSION}
\renewcommand{\theequation}{A\arabic{equation}}
\setcounter{equation}{0}

In this appendix we calculate the second and fourth cumulants of a single random telegraph noise for the CPMG sequence and confirm the convergence of our exact results for pure dephasing with the cumulant expansion in the weak coupling limit, $\eta=v/\gamma \ll 1$, where non-Gaussian effects are reduced.\cite{Cywinski} Including the first two non-vanishing cumulants, we have
\begin{equation}
\chi (t) \approx e^{-\frac{C_2(t)}{2}+\frac{C_4(t)}{24}},
\label{chicum}
\end{equation}
where
\begin{equation}
C_2(t)=\langle \Phi^2 (t) \rangle, \hspace{0.4 cm} C_4(t)=\langle \Phi^4 (t) \rangle -3 \langle \Phi^2 (t) \rangle^2.
\label{cum}
\end{equation}
$\langle \Phi^n (t) \rangle$ is the $n$-th moment given by:
\begin{eqnarray}
\langle \Phi^n (t) \rangle \!&\!=\!&\! n!\, v^n \int_0^t dt_1 \int_0^{t_1} dt_2 \cdots \int_0^{t_{n-1}} dt_n f(t_1) \cdots \nonumber \\ &&  f(t_n) e^{-2\gamma [t_1-t_2+ \cdots -(-1)^n t_n]},
\end{eqnarray}
where $f(t)$ is the pulse sequence function.\cite{Cywinski} In the Gaussian approximation all terms higher than $C_2$ vanish, thus the ratio between $C_4$ and $C_2$ provides a measure for the  non-Gaussianity of the RTN. Calculating the cumulants, Eq.~(\ref{cum}), for a small number of pulses, one obtains a pattern that allows a full summation within each cumulant order leading to general $N$ formulas. Denoting $x\equiv \gamma \tau$ we find for a symmetric TLF:
\begin{eqnarray}
C_2(t) \!\!&\!\!=\!&\!\! -\frac{\eta^2}{2} \left[ \left( 1-\!(-1)^N e^{-2 \gamma t} \right) (1- {\rm sech} x )^2 + \right. \nonumber \\ && \left. 2N(\tanh x -x) \right]
\end{eqnarray}
and
\begin{eqnarray}
C_4(t) \!\!&\!\!=\!&\!\! \frac{3 \eta^4}{4} \! \left\{ (1\!-\! e^{-4 \gamma t})({\rm sech} x \!-\! 1)^4\!\!+4(1-\!(-1)^N e^{-2\gamma t} ) \right. \nonumber \\ && \left. \times \left[ \tanh^4 x -2x \tanh x \, {\rm sech} x (1-{\rm sech} x) \right] + \right. \nonumber \\ && \left. 4N \left[ 2(1+e^{-2 \gamma t}) (\tanh x -x)(1- {\rm sech}x)^2 + \right. \right. \nonumber \\ && \left. \left. (\tanh x -x)(4{\rm sech} x-1)+\tanh^3 x \right] \right\}.
\label{C24}
\end{eqnarray}
These expressions reduce to the fourth cumulant results for SE and 2-pulse CPMG reported in Ref.~\onlinecite{Cywinski}.\cite{doublexi}

Expanding Eq.~(\ref{lsym}) to fourth order in $\eta$, the eigenvalues are:
\begin{eqnarray}
\lambda_\pm \!&\!=\!&\! \pm e^{\pm x} -\frac{\eta^2}{2} \left( x e^{\pm x} \pm {\rm sech} x \right)+ \frac{\eta^4}{8} \left[ \pm e^{\pm x} x (x \mp 1) \mp \right. \nonumber \\
&& \left. {\rm sech} x ( {\rm sech}^2 x +2x \tanh x ) \right] +\mathcal{O} (\eta^6),
\end{eqnarray}
and the expansion of $\chi_{CP}(t)$ in Eq.~(\ref{depCPs}) matches Eqs.~(\ref{chicum}), and (\ref{C24}) to fourth order in $\eta$. We note that while there seem to be differences in the cumulant expressions between even and odd number of pulses, these differences cancel out when the cumulants are summed to infinite order to provide the exact result of Eq.~(\ref{depCPs}).

\section*{APPENDIX B: TWO FLUCTUATORS AT PURE DEPHASING}
\renewcommand{\theequation}{B\arabic{equation}}
\setcounter{equation}{0}

In this appendix we show how to extend the single TLF theory to two or more TLFs for the case of pure dephasing. Considering two symmetric TLFs, we split the probability to accumulate a phase $\phi$ at time $t$ into four partial probabilities, corresponding to the possible combinations of two TLF states at that time:
\begin{equation*}
p(\phi,t)\!=p_{++}(\phi,t)\!+p_{+-}(\phi,t)\!+p_{-+}(\phi,t)\!+p_{--}(\phi,t).
\end{equation*}
Taking a short time increment $\tau$, during which the switching probabilities are $\gamma_i \tau$, we have
\begin{eqnarray}
p_{++}(\phi ,t \!&\!+\!&\! \tau )\! = (1\!-\!\gamma _1 \tau )(1\!-\! \gamma _2\tau )p_{++}(\phi \!-\! v_1\tau \!-\! v_2\tau ,t) + \nonumber \\ && (1 -\gamma _1\tau )(\gamma _2\tau) p_{+-}(\phi - v_1\tau + v_2\tau ,t) + \nonumber \\ && (\gamma _1 \tau)(1 - \gamma _2\tau ) p_{-+}(\phi + v_1\tau - v_2\tau ,t) + \nonumber \\
&& (\gamma _1 \tau) (\gamma _2 \tau) p_{--}(\phi + v_1\tau + v_2\tau ,t) + \nonumber \\ && \int\limits_t^{t + \tau } d{t_1} \left[ p_{-+}\left( {\phi + {v_1}(t + \tau - t_1),t_1} \right) + \right. \nonumber \\ && \hspace{1.1 cm}  \left. p_{+-}\left( {\phi + v_2(t + \tau -t_1),t_1} \right) \right],
\label{ppp}
\end{eqnarray}
and similar equations for the other three partial probabilities. The four rate equations are found by taking infinitesimal $\tau$ and keeping only linear terms in $\tau$:
\begin{eqnarray}
\!\!\!\!\!\!\!\! \dot{p}_{++}\!\!&\!\!=\!&\!\! -(\gamma_1\!\!+\!\gamma_2)p_{++}\!\!+\!\gamma_1 p_{-+}\!\! +\!\gamma_2 p_{+-}\!\!-\!(v_1\!\!+\!v_2) \partial_\phi p_{++} \nonumber \\
\!\!\!\!\!\!\!\! \dot{p}_{+-}\!\!&\!\!=\!&\!\! -(\gamma_1\!\!+\!\gamma_2)p_{+-}\!\!+\!\gamma_1 p_{--} \!\!+\!\gamma_2 p_{++}\!\!-\!(v_1\!\!-\!v_2) \partial_\phi p_{+-} \nonumber \\
\!\!\!\!\!\!\!\! \dot{p}_{-+}\!\!&\!\!=\!&\!\! -(\gamma_1\!\!+\!\gamma_2)p_{-+}\!\!+\!\gamma_1 p_{++} \!\!+\!\gamma_2 p_{--}\!\!+\!(v_1\!\!-\!v_2) \partial_\phi p_{-+} \nonumber \\
\!\!\!\!\!\!\!\! \dot{p}_{--}\!\!&\!\!=\!&\!\! -(\gamma_1\!\!+\!\gamma_2)p_{--}\!\!+\!\gamma_1 p_{+-}\!\!+\! \gamma_2 p_{-+}\!\!+\!(v_1\!\!+\!v_2) \partial_\phi p_{--}.
\label{pij}
\end{eqnarray}
Denoting $\chi_{ij}(t)$ as the  phase factors averaged over switching histories ending at TLF states $(i,j)$, where $(i,j)=\{+,-\}$, we construct four combinations of phase factors:
\begin{eqnarray}
\chi (t)\!&\!=\!&\! \chi_{++}(t)+\chi_{+-}(t)+\chi_{-+}(t)+\chi_{--}(t) \nonumber \\
\chi _1(t)\!&\!=\!&\! \chi_{++}(t)+\chi_{+-}(t)-\chi_{-+}(t)-\chi_{--}(t) \nonumber\\
\chi _2(t)\!&\!=\!&\!\chi_{++}(t)-\chi_{+-}(t)+\chi_{-+}(t)-\chi_{--}(t)\nonumber \\
\chi _3(t)\!&\!=\!&\! \chi_{++}(t)-\chi_{+-}(t)-\chi_{-+}(t)+\chi_{--}(t),
\label{chi14}
\end{eqnarray}
where $\chi(t)$ corresponds to the qubit decay and the other three $\chi_i(t)$ generalize $\delta \chi$, introduced below Eq.~(\ref{chifirst}), to the two-TLF case. The rate equations, Eqs.~(\ref{pij}), are translated to a set of coupled equations, analogous to Eq.~(\ref{M1}):
\begin{equation}
\left( \!{\begin{array}{*{40}{c}}
{\dot \chi }\\
\dot{\chi_1} \\ \dot{\chi_2} \\ \dot{\chi_3}
\end{array}} \right) = {M_1} \! \left(\! {\begin{array}{*{40}{c}}
{\chi }\\
\chi_1 \\ \chi_2 \\ \chi_3
\end{array}} \right),
\end{equation}
where $M_1$ is given by
\begin{equation}
M_1 = \left( {\begin{array}{*{20}{c}}
0&{ - i{v_1}}&{ - i{v_2}}&0\\
{ - i{v_1}}&{ - 2{\gamma _1}}&0&{ - i{v_2}}\\
{ - i{v_2}}&0&{ - 2{\gamma _2}}&{ - i{v_1}}\\
0&{ - i{v_1}}&{ - i{v_2}}&{ - 2\left( {{\gamma _1} + {\gamma _2}} \right)}
\end{array}} \right).
\end{equation}
After a $\pi$ pulse, the qubit evolves under $M_2$, defined by substituting $v_i \rightarrow -v_i$ in $M_1$. As in the single-TLF case, the qubit signal under $N$-pulse sequence is calculated by diagonalizing the evolution operator $T$, defined by Eq.~(\ref{T}), with $L={\rm diag} (-1,1,1,-1)$. The eigenvalues of $T$ are found to factor out for the two fluctuators, as stated in Eq.~(\ref{chint}). For $n_T$ fluctuators the evolution operator is a square matrix of size $2^{n_T}$ and it can be shown by induction that Eq.~(\ref{chint}) holds for any number of fluctuators.

\bibliographystyle{amsplain}

\begin{thebibliography}{99}
\bibitem{rotated} Notice that while the PDD rates are unchanged from those calculated in Ref.~\onlinecite{Ramon2} for a rotated reference frame, the CPMG rates, reported here, correct previous results.
\bibitem{delta} As explained in Ref.~\onlinecite{Ramon2}, the $2 \gamma$ decay rates are predominantly associated with the three variables ($\delta x, \delta y, \delta z$) that are coupled to the Bloch vector components in the six differential equations.
\bibitem{3TLF} We have worked out the set of 12 coupled equations for the case of three TLFs at the OP, excplicitly showing that the general solution given in Eq.~(\ref{PDDn}) is correct.
\bibitem{StrongnT} Using our full analytical solutions for $n_T=2$ and $n_T=3$ as starting points, we were able to obtain by induction approximate analytical solutions that work for a limited range of $n_T$ values (up to $n_T \sim 10$). These solutions contain two decay rates, which have distinct structure for even and odd number of TLFs (not reported here).
\bibitem{doublexi} Note that we define the RTN noise with $\xi (t)= \pm 1$ as opposed to $\xi (t)= \pm 1/2$ in Ref.~\onlinecite{Cywinski}, which results in doubling the coupling strength and $\eta$ in our formulae.
\end{thebibliography}

\end{document}